\documentclass[useAMS,usenatbib]{mnras}

\usepackage[dvipsnames]{xcolor}
\usepackage{tikz,hyperref}
\usepackage{ulem}

\definecolor{lime}{HTML}{A6CE39}
\DeclareRobustCommand{\orcidicon}{
	\begin{tikzpicture}
	\draw[lime, fill=lime] (0,0) 
	circle [radius=0.13] 
	node[white] {{\fontfamily{qag}\selectfont \tiny ID}};
	\draw[white, fill=white] (-0.0625,0.095) 
	circle [radius=0.007];
	\end{tikzpicture}
	\hspace{-2mm}
}

\foreach \x in {A, ..., Z}{\expandafter\xdef\csname orcid\x\endcsname{\noexpand\href{https://orcid.org/\csname orcidauthor\x\endcsname}
			{\noexpand\orcidicon}}
}

\usepackage{graphicx}
\usepackage{longtable}
\usepackage{amssymb,amsmath}

\newcommand{\heiiwr}{He\,{\sc ii}\,$\lambda4686$}
\newcommand{\heiibyhb}{He\,{\sc ii}\,$\lambda4686$/H$\beta$}
\newcommand{\ewhb}{EW(H$\beta$)}

\newcommand{\msol}{M$_{\odot}$}

\newcommand{\hii}{H\,{\small{\sc ii}}}
\newcommand{\ha}{H$\alpha$}
\newcommand{\hb}{H$\beta$}

\newcommand{\heii}{He\,{\sc ii}}
\newcommand{\oi}{[O\,{\sc i}]\,$\lambda6300$}
\newcommand{\oibyha}{[O\,{\sc i}]\,$\lambda6300$/H$\alpha$}
\newcommand{\oiii}{[O\,{\sc iii}]\,$\lambda5007$}

\newcommand{\nii}{[N\,{\sc ii}]}

\newcommand{\ariv}{[Ar\,{\sc iv}]}
\newcommand{\arivbyariii}{[Ar\,{\sc iv}]\,$\lambda4711+4740$/[Ar\,{\sc iii}]\,$\lambda7135$}

\newcommand{\oiiibyhb}{[O\,{\sc iii}]\,$\lambda5007$/H$\beta$}
\newcommand{\niibyha}{[N\,{\sc ii}]\,$\lambda6583$/H$\alpha$}
\newcommand{\siibyha}{[S\,{\sc ii}]\,$\lambda6717+6731$/H$\alpha$}
\newcommand{\oibyoiii}{[O\,{\sc i}]\,$\lambda6300$/[O\,{\sc iii}]\,$\lambda5007$}
\newcommand{\oiiibyoii}{[O\,{\sc iii}]\,$\lambda5007$/[O\,{\sc ii}]\,$\lambda7325$}

\newcommand{\ergs}{erg\,s$^{-1}$}
\newcommand{\ergcms}{erg\,cm$^{-2}$\,s$^{-1}$}

\newcommand{\ergcmsarc}{erg\,cm$^{-2}$\,s$^{-1}$\,arcsec$^{-2}$}
\newcommand{\logU}{$\log\langle U\rangle$}

\title[HeII in the Cartwheel galaxy]{
Detection of He$^{++}$ ion in the star-forming ring of the Cartwheel using MUSE data and ionizing mechanisms
}
\author[Y.\,D.\,Mayya et al.]{Y.\,D.\,Mayya\thanks{Email: ydm@inaoep.mx}$^{1\orcidA{}}$, 
A. Plat$^{2\orcidB{}}$, V.\,M.\,A. G\'omez-Gonz\'alez$^{3\orcidC{}}$, J. Zaragoza-Cardiel$^{1,4\orcidD{}}$, 
\newauthor
S. Charlot$^{5\orcidE{}}$ and G. Bruzual$^{6\orcidF{}}$
\\
$^{1}$Instituto Nacional de Astrof{\'\i}sica, \'Optica y Electr\'onica, Luis Enrique Erro 1, Tonantzintla 72840, Puebla, Mexico\\
$^{2}$Steward Observatory, 933 N. Cherry Avenue, University of Arizona, Tucson, AZ 85721, USA \\
$^{3}$Institute for Physics and Astronomy, Universit\"{a}t Potsdam, Karl-Liebknecht-Str. 24/25, D-14476 Potsdam, Germany\\
$^{4}$Consejo Nacional de Ciencia y Tecnolog\'ia, Av. Insurgentes Sur 1582, 03940,  Mexico City, Mexico\\
$^{5}$Sorbonne Universit\'e, CNRS, UMR7095, Institut d'Astrophysique de Paris, F-75014, Paris, France \\
$^{6}$Instituto de Radioastronom\'{i}a y Astrof\'{i}sica, UNAM Campus Morelia, Apartado postal 3-72, 58090 Morelia, Michoac\'{a}n, Mexico\\
}

\begin{document}
\maketitle

\begin{abstract}
We here report the detection of the nebular \heiiwr\ line in 32 \hii\ regions in the metal-poor collisional ring galaxy Cartwheel using the Multi-Unit Spectroscopic Explorer (MUSE) dataset. The measured I(\heiiwr)/I(\hb) ratio varies from 0.004 to 0.07, with a mean value of 0.010$\pm$0.003. Ten of these 32 \hii\ regions are coincident with the location of an Ultra Luminous X-ray (ULX) source. We used the flux ratios of important diagnostic lines and results of photoionization by Simple Stellar Populations (SSPs) to investigate the likely physical mechanisms responsible for the ionization of He$^+$. We find that the majority of the regions (27) are consistent with photoionization by star clusters in their Wolf-Rayet (WR) phase with initial ionization parameter $-3.5<$\logU$< -2.0$. Blue Bump (BB), the characteristic feature of the WR stars, however, is not detected in any of the spectra. We demonstrate that this non-detection is due to the relatively low equivalent width (EW) of the BB in metal-poor SSPs, in spite of  containing sufficient number of WR stars to reproduce the observed I(\heiiwr)/I(\hb) ratio of $\le$1.5\% at the Cartwheel metallicity of Z=0.004. The \hii\ regions in the WR phase that are coincident with a ULX source do not show line ratios characteristic of ionization by X-ray sources. However, the ULX sources may have a role to play in the ionization of He$^+$ in two (\#99, 144) of the five regions that are not in the WR phase. Ionization by radiative shocks along with the presence of channels for the selective leakage of  ionizing photons are the likely scenarios in \#17 and \#148, the two regions with the highest observed I(\heiiwr)/I(\hb) ratio.
\end{abstract}


\section{Introduction}

Ionization of He$^+$ requires environments that generate energies in excess of 
54.4\,eV. Detection of the \heiiwr\ nebular line in star-forming galaxies 
suggests that processes related to star formation are able to create
such environments. Photoionization by hard ultraviolet photons emanating
from Wolf--Rayet (WR) stars is the front-runner among these processes
\citep{Schaerer1996}. 
Simple stellar population (SSP) models incorporating the state-of-the-art 
developments in massive star evolution and modeling of the wind-dominated
atmospheres of O-type and WR stars are able to explain the observed I(\heiiwr)/I(\hb)
intensity ratios (\heiibyhb\ for short) in at least some star-forming galaxies \citep{Plat2019}.
These models predict a maximum value of \heiibyhb=0.01 during 
the WR phase ($\sim$3--5~Myr) for metallicities $Z\ge$0.004, with a decrease
of this ratio at lower metallicities \citep[see][and references therein]{Mayya2020}.
The observed values of the \heiibyhb\ ratio in metal-poor galaxies often are found to
be above 0.01 \citep{Shirazi2012, Kehrig2015, Kehrig2018, Schaerer2019}. 
The SSP models of \citet{Eldridge2017} involving binary stars help to alleviate the 
problem to some extent, but explaining the presence of He$^{++}$ ion in high \hb\ 
equivalent width (EW) systems (burst ages $<$3~Myr) remains a challenge. 
This has forced exploration of alternative mechanisms of ionization.
\citet{Plat2019} carried out an exhaustive analysis of the physical
mechanisms that can produce the observed ratio of \heiibyhb. They find the
need for one or more of the following processes at work in order to produce
the observed high values: 
(1) the presence of stars significantly more massive than 100~\msol;
(2) extremely high ionization parameter, $\log(U)>-1$;
(3) hard radiation from binary stars (in particular, X-ray binaries);
(4) ionization of He$^+$ by radiative shocks; and/or
(5) ionization of He$^+$ by an active galactic nucleus (AGN), if and when present.

Over the last decade or so, it has been possible to obtain reliable fluxes
of the relatively weak \heiiwr\ line for a large sample of objects, thanks to 
dedicated spectroscopic surveys such as Sloan Digital Sky Survey (SDSS). 
Given that the often-used nebular diagnostic lines 
\citep[see e.g.][]{Perez-Montero2017}
are much brighter than the \heiiwr\ line, the available dataset has also 
allowed accurate determination of nebular metallic abundances.
Dataset obtained from these large surveys are the ones often used to confront 
with the predictions of SSP models \citep[e.g.][]{Plat2019, Schaerer2019}.
These data typically sample zones that span several kiloparsecs in size,
centered on the nucleus of the galaxy. Though the diagnostic lines \citep{BPT1981}
allow the rejection of spectra dominated by an AGN at least at high enough metallicities 
\citep[$Z>$0.008; see][]{Feltre2016},
the presence of a weak AGN in a spectrum dominated by a starburst component 
cannot be ruled out. Over the large spatial scales sampled in these studies, 
several of the above mentioned five physical mechanisms are likely to be at 
work, thus inhibiting discerning the 
relative importance of each mechanism. Data on smaller scales, ideally of 
selected regions in nearby galaxies, are required so as to explore the role
of each of the above-listed mechanisms in increasing the \heiibyhb\ ratio above the 
canonical values predicted by the SSP models. 

Availability of spectrographs incorporating integral field units (IFUs) on
large telescopes such as Multi Unit Spectroscopic Explorer (MUSE) on the 
Very Large Telescope (VLT) \citep{Bacon2010}
and MEGARA on the Gran Telescopio Canarias (GTC) \citep{2018SPIE10702E..17G} 
is allowing such studies possible in recent years. 
\citet{Kehrig2015} and \citet{Kehrig2018} used MUSE data to spatially
map the \heiiwr\ line in
two of the most metal-poor galaxies: I\,Zw\,18 and SBS\,0335~$-$~052E, finding that the 
observed He$^+$ ionization cannot be explained by the WR stars present in 
these galaxies. Recently, \citet{Mayya2020} used MEGARA to map the central 
starburst cluster of NGC\,1569, a galaxy with the oxygen abundance similar to 
that of the Large Magellanic Cloud (LMC), finding that the WR stars in the starburst cluster
are able to completely explain the observed ionization. 
Data on regions where hydrogen ionization is dominated by photons
from massive stars of metallicity below that of the LMC are needed to address
the sources of He$^+$ ionization in distant metal-poor galaxies. The collisional-ring galaxy Cartwheel 
provides such a laboratory, as explained below.

\begin{figure*}
\begin{centering}
\includegraphics[width=1.0\linewidth]{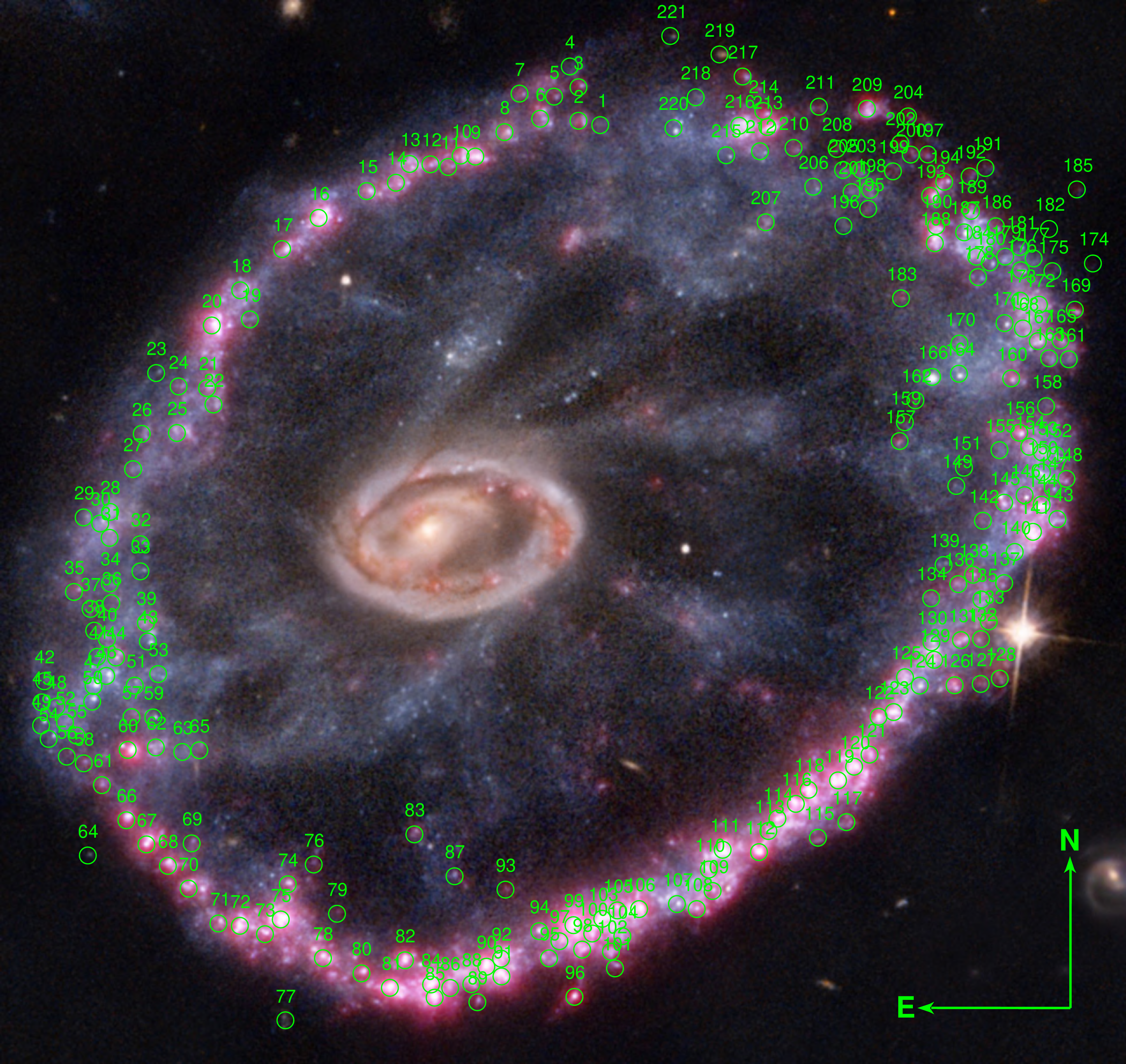}
\par\end{centering}
\caption{Colour-composite HST/WFPC2 and VLT/MUSE image of the Cartwheel galaxy. The RGB image is formed using the HST/WFPC2 filters (PSF=0.2~arcsec) in F814W, pseudo-green and F450W as red, green and blue components, respectively; the \ha\ image is constructed from MUSE data (PSF=0.6~arcsec) as a fourth reddish component. The \hii\ regions in the ring are identified by numbers and shown by green circles of 0.6~arcsec radius, which corresponds to 370~pc at the distance of the Cartwheel.
}
\label{fig:muse_image}
\end{figure*}

The Cartwheel is considered as the archetype of the class of collisional ring galaxies
\citep{1996FCPh...16..111A, 2010MNRAS.403.1516S}. Ring galaxies
are characterized by a ring that harbours a chain of star forming knots.
The star formation in the ring is believed to be triggered by a radially expanding
density wave that was formed as a result of a compact galaxy 
plunging through a massive gas-rich disk galaxy close to its center and almost
perpendicular to it \citep{1976ApJ...209..382L}.
\citet{Higdon1995} found that the \ha\ emission in the Cartwheel, a tracer 
of current star formation, is distributed along a ring as predicted by the
collisional scenario of the formation of ring galaxies. 
MUSE dataset is available on this galaxy at the
seeing-limited spatial resolution of $\sim$0.6~arcsec. This dataset provides 
optical spectra covering a rest wavelength range of $\sim$4600 to 9100~\AA\ over the entire galaxy.
On the \ha\ image constructed using this dataset, we have identified more
than 200 individual \hii\ regions in and around the ring. A colour-composite image
formed using this \ha\ image is shown in Figure~\ref{fig:muse_image}. 
At the distance of the Cartwheel (128~Mpc using the Hubble constant of 
71~km\,s$^{-1}$\,Mpc$^{-1}$), MUSE spectra are available at physical 
scales of $\sim$370~pc, which is an order of magnitude better as compared 
to the typical size scale where \heiiwr\ is detected in metal-poor galaxies.
At the spatial resolution of the Wide Field and Planetary Camera 2 (WFPC2) images of 
the Hubble Space Telescope (HST), which are the highest resolution images 
($\sim$0.2~arcsec=125~pc) available for this galaxy, we can associate 
each MUSE-identified \hii\ region with a population of super star clusters 
(SSCs), which provide the ionization of the \hii\ regions.
\citet{Fosbury1977} have measured an oxygen abundance of 12+log(O/H)$\sim$8.0, corresponding roughly to $Z\sim0.003$ \citep[see Table 2 of][]{Gutkin2016},
a value at which the observed \heiibyhb\ ratio in galaxy samples is higher than that predicted by most of the SSPs. The wide spectral coverage of MUSE data allowed the flux measurement of nebular lines useful to study the ionization state of the \hii\ regions using the standard line ratio diagrams  \citep[][hereafter BPT]{BPT1981}.
We used this new dataset to measure an average oxygen abundance for the ring regions of 12+log(O/H)$\sim$8.19$\pm$0.15 \citep{Zaragoza2022}, which is marginally higher than the value reported by \citet{Fosbury1977} for three of the brightest \hii\ regions.

An additional aspect that makes the Cartwheel a good candidate for 
understanding the \heii\ ionization problem is the presence of more than
15 ultra-luminous X-ray sources (ULXs) in its star-forming ring  
\citep{Gao2003, Wolter2004}. In fact the Cartwheel is the record 
holder for the maximum number of ULX sources in a single galaxy \citep{Wolter2018}.
The ULX emission is believed to be originating
in high-mass X-ray binaries (HMXBs), with the compact object most likely an
intermediate-mass black hole (IMBH) \citep{Mapelli2010, Wolter2018}. 
The presence of these sources allows us to explore the role of HMXBs in the
He$^+$ ionization in regions where massive stars contribute to the ionization of
hydrogen and other ions of ionization potential much lower than 54.4\,eV.

In Section~2, we describe the dataset, extraction of individual spectrum, and 
details of measurement of line fluxes. 
The analysis of nebular line ratios is described in Section 3.
A detailed discussion of results on the He$^+$ ionization in each star-forming complex is carried out in Section~4. Our conclusions are given in Section~5.

\begin{table*}
\small\addtolength{\tabcolsep}{-3pt}
\caption{Cartwheel \hii\ regions with \heiiwr\ nebular emission line.}\label{tab:heii_regions}
\begin{tabular}{lllcrrrccrrrrrrrl}
\hline
\multicolumn{3}{c}{ID} && \multicolumn{4}{c}{\hb} && \multicolumn{2}{c}{\heii} && \multicolumn{4}{c}O and WR stars & Cont\\
 \cline{1-3}
 \cline{5-8}
 \cline{10-11}
 \cline{13-16}
\hii\ & Higdon & ULX && log f(\hb)      & SNR   & EW     &log M$_\ast$ & & log Q(He$^+$) & I(\heii)/I(\hb) & & $N_{\rm O7V}$ & EW$_{\rm BB}$ & $N_{\rm WNL}$ & $N_{\rm WR}$ & SNR  \\
     &      &         && [\ergcms]         &       & [\AA]  & \msol\ & &ph\,s$^{-1}$ & $\times$100 & & & 3$\sigma$ & C\&B & C\&B & \\ 
 (1) & (2)  & (3)     &&  (4)              & (5)   &   (6)   &  (7)  & &  (8)   &      (9)        & & (10)  & (11)  & (12)& (13)& (14)\\  
\hline
 10  &  H1  & ---     &&$-15.082\pm0.009$  & 109.9 &   57.2  & 4.84 & & 49.29 & $1.186\pm0.326$  & &   341 &   6.07 &   2 &   7 & 17.9\\ 
 16  &  H3  & G18     &&$-14.672\pm0.005$  & 183.5 &   71.7  & 5.25 & & 49.56 & $0.845\pm0.196$  & &   876 &   3.28 &   7 &  19 & 34.1\\ 
 17  &  H4  & ---     &&$-15.533\pm0.014$  &  70.7 &   29.6  & 4.39 & & 49.19 & $2.666\pm0.784$  & &   120 &   5.07 &   1 &   2 & 21.9\\ 
 20  &  H5  & ---     &&$-14.841\pm0.005$  & 204.6 &   63.7  & 5.08 & & 49.55 & $1.232\pm0.168$  & &   594 &   2.77 &   4 &  13 & 38.3\\ 
 60  &  H6  & ---     &&$-14.808\pm0.006$  & 165.3 &  154.4  & 5.11 & & 49.43 & $0.861\pm0.160$  & &   640 &   7.22 &   5 &  14 & 14.7\\ 
 75  &  H10 & G19/W23 &&$-14.512\pm0.003$  & 287.3 &   90.5  & 5.41 & & 49.68 & $0.775\pm0.151$  & &  1267 &   3.48 &  10 &  28 & 34.0\\ 
 81  &  H12 & ---     &&$-14.366\pm0.003$  & 363.8 &  159.6  & 5.55 & & 50.03 & $1.256\pm0.098$  & &  1773 &   3.43 &  14 &  39 & 30.2\\ 
 84  &  H14 & ---     &&$-14.364\pm0.002$  & 408.3 &  120.7  & 5.56 & & 49.91 & $0.936\pm0.110$  & &  1781 &   2.93 &  14 &  39 & 35.4\\ 
 85  &  H14 & G17/W7  &&$-14.517\pm0.003$  & 303.4 &   79.8  & 5.40 & & 49.87 & $1.221\pm0.149$  & &  1252 &   2.66 &  10 &  27 & 38.3\\ 
 86  &  H14 & ---     &&$-14.994\pm0.005$  & 195.7 &   77.8  & 4.93 & & 49.12 & $0.648\pm0.195$  & &   417 &   4.91 &   3 &   9 & 21.7\\ 
 90  &  H15 & ---     &&$-14.339\pm0.003$  & 338.4 &  126.8  & 5.58 & & 49.80 & $0.686\pm0.101$  & &  1887 &   3.04 &  15 &  41 & 33.2\\ 
 91  &  H15 & G14     &&$-14.756\pm0.006$  & 170.9 &   49.4  & 5.16 & & 49.55 & $1.008\pm0.207$  & &   722 &   2.14 &   6 &  15 & 46.0\\ 
 92  &  H15 & ---     &&$-14.808\pm0.005$  & 194.1 &   66.4  & 5.11 & & 49.42 & $0.854\pm0.195$  & &   640 &   3.15 &   5 &  14 & 31.0\\ 
 97  &  H17 & ---     &&$-14.984\pm0.007$  & 136.5 &   68.8  & 4.94 & & 49.18 & $0.733\pm0.161$  & &   427 &   3.13 &   3 &   9 & 35.6\\ 
 98  &  H17 & ---     &&$-14.887\pm0.006$  & 163.3 &   72.9  & 5.03 & & 49.17 & $0.565\pm0.171$  & &   534 &   3.70 &   4 &  11 & 27.6\\ 
 99  &  H17 & G12/W24 &&$-13.575\pm0.001$  &1055.0 &  282.6  & 6.35 & & 50.55 & $0.664\pm0.052$  & & 10960 &   2.78 &  92 & 242 & 35.2\\ 
100  &  H17 & ---     &&$-14.721\pm0.005$  & 221.9 &   69.9  & 5.20 & & 49.42 & $0.694\pm0.159$  & &   783 &   2.68 &   6 &  17 & 39.2\\ 
105  &  H19 & ---     &&$-14.824\pm0.006$  & 171.0 &   61.4  & 5.10 & & 49.31 & $0.676\pm0.162$  & &   617 &   2.63 &   5 &  13 & 42.0\\ 
106  &  H19 & ---     &&$-14.563\pm0.004$  & 241.5 &   81.3  & 5.36 & & 49.57 & $0.683\pm0.161$  & &  1126 &   2.99 &   9 &  24 & 34.2\\ 
108  &  H21 & ---     &&$-15.148\pm0.009$  & 116.2 &   52.5  & 4.77 & & 49.32 & $1.461\pm0.323$  & &   292 &   4.66 &   2 &   6 & 21.8\\ 
111  &  H20 & G11/W10 &&$-14.894\pm0.007$  & 152.5 &   34.9  & 5.03 & & 49.43 & $1.066\pm0.240$  & &   525 &   2.38 &   4 &  11 & 47.5\\
112  &  H22 & ---     &&$-14.493\pm0.004$  & 271.0 &  219.4  & 5.43 & & 49.37 & $0.361\pm0.107$  & &  1323 &   5.50 &  11 &  29 & 22.5\\ 
113  &  H22 & ---     &&$-15.201\pm0.009$  & 109.4 &   46.1  & 4.72 & & 49.02 & $0.831\pm0.213$  & &   259 &   4.07 &   2 &   5 & 26.4\\ 
116  &  H23 & ---     &&$-14.614\pm0.004$  & 259.1 &   93.4  & 5.31 & & 49.81 & $1.341\pm0.153$  & &  1001 &   3.14 &   8 &  22 & 34.1\\ 
118  &  H23 & G8      &&$-14.432\pm0.003$  & 360.0 &  123.3  & 5.49 & & 49.71 & $0.686\pm0.126$  & &  1523 &   3.24 &  12 &  33 & 35.1\\ 
119  &  H24 & G6/W14  &&$-14.498\pm0.006$  & 170.1 &   60.6  & 5.42 & & 49.83 & $1.053\pm0.184$  & &  1308 &   2.61 &  11 &  28 & 39.1\\ 
120  &  H24 & ---     &&$-14.679\pm0.007$  & 152.4 &   51.0  & 5.24 & & 49.60 & $0.951\pm0.251$  & &   862 &   2.98 &   7 &  19 & 36.5\\ 
121  &  H24 & ---     &&$-14.983\pm0.011$  &  91.3 &   69.8  & 4.94 & & 49.28 & $0.908\pm0.239$  & &   428 &   5.69 &   3 &   9 & 18.4\\ 
141  &  H26 & ---     &&$-14.700\pm0.005$  & 188.6 &   37.4  & 5.22 & & 49.56 & $0.918\pm0.223$  & &   821 &   2.11 &   6 &  18 & 54.6\\ 
144  &  H26 & G2/W17  &&$-15.036\pm0.007$  & 134.8 &   38.0  & 4.88 & & 49.61 & $2.223\pm0.322$  & &   379 &   3.01 &   3 &   8 & 34.8\\ 
148  &  --- & ---     &&$-15.785\pm0.021$  &  46.8 &   21.8  & 4.14 & & 49.31 & $6.283\pm1.599$  & &    67 &   7.90 &   0 &   1 & 13.8\\ 
213  &  H29 & G9/W12  &&$-14.907\pm0.006$  & 164.5 &   88.9  & 5.01 & & 49.20 & $0.644\pm0.196$  & &   510 &   4.53 &   4 &  11 & 24.4\\ 
\hline
\end{tabular}
(1) \hii\ region identification number used in this work, (2) \hii\ complex number from \citet{Higdon1995}, (3) ULX sources associated to the \hii\ region, (4-7) Measured and derived quantities using \hb\ nebular line, where stellar mass in column 7 is derived assuming a Lyman continuum photons rate of 4.94$\times10^{46}$ using C\&B model corresponding to an age of 3.4~Myr for the ionizing cluster, (8) He$^+$ ionizing photon rate determined from the observed \heiiwr\ flux for a Case~B \hii\ region, (9) 100$\times$ the intensity of \heiiwr\ line with respect to that of \hb, (10) Equivalent number of O7V stars derived assuming a typical \hb\ luminosity of 4.76$\times10^{36}$~\ergs\ for an O7V star from \citet{Esteban2010}, (11) Upper limit for the equivalent width of the BB from WNL type inferred as the 3$\sigma$ flux of a broad (FWHM=20~\AA) BB divided by the underlying continuum flux, (12-13) The number of WNL and total WR stars expected in C\&B models during the WR phase for the cluster mass in column~7, (14) SNR of the continuum adjacent to the BB bump.
\end{table*}

\section{The sample of He$^{++}$ nebulae and the control sample of bright \hii\ regions}

\subsection{The \hii\ region parent sample}

Given that the ionization potential of hydrogen is four times lower as compared to the second
ionization potential of helium, the He$^{++}$ nebulae are expected to be a subset of the ionized nebulae.
The outer ring of the Cartwheel is currently experiencing an intense burst of star formation. 
\citet{Higdon1995} found that the \ha\ emission, a tracer of current star formation, 
is predominantly confined to 29 ionized complexes in the outer ring. 
As a first step, we used the MUSE datacube to obtain a narrow-band image at the 
red-shifted wavelength of the \ha\ line to locate ionized regions at the 
resolution of MUSE (FWHM=0.6~arcsec$\sim$370~pc),
which is $\sim$3 times better as compared to the image used by \citet{Higdon1995}.
In Figure~\ref{fig:muse_image}, we show an RGB image formed by combining
the HST/WFPC2 filters in F814W, pseudo-green and F450W as red, green and blue components, respectively;
the \ha\ image constructed from MUSE data is used as a fourth reddish component.
We identified visually 221 prominent \ha-emitting knots in the ring or near to it on the MUSE image, which are shown by numbers in this image. At the resolution of the HST/WFPC2 images, most of the \ha-emitting knots are associated with one or more compact clusters, which are the most likely ionizing sources of the nebulae. Thus, the selected regions are star-forming complexes spread over the entire extracted area. However, at the spatial resolution offered by MUSE, each complex is basically a compact unresolved knot. We hence used a uniform aperture of 0.6~arcsec radius to extract spectrum of each region. 
We note that the \ha\ surface brightness in the outer ring is 10 to 100 times brighter than the typical cut-off surface brightness of 2$\times10^{-17}$~\ergcmsarc\ for the Diffuse Ionized Gas (DIG) \citep[see e.g.][]{Belfiore2022}, and hence the DIG contribution to the flux ratios of lines in the extracted spectra can be ignored.

We did not attempt to subtract the disk spectrum from the extracted spectra as it is non-trivial to locate a region for extraction of the local disk spectrum in a galaxy such as the Cartwheel because of its peculiar morphology and star formation history. Thus the extracted spectra contain contributions from any pre-collisional, as well as from all post-collisional, populations in the extracted area. From an analysis of stellar populations in the extracted spectra, \citet{Zaragoza2022} concluded that the spectra lack absorption features characteristics of disk stellar populations. This is understandable because the pre-collsional disk is around a factor of five fainter than the ring \citep{Marcum1992, Higdon1995}, and hence the extracted spectra are dominated by contribution from all populations formed after the collision around 100~Myr ago \citep{Renaud2018}. The extracted spectra, however, could contain contribution from non-ionizing populations formed over the last 100~Myr. Such a population will not contribute to the measured emission line fluxes, but will make the observed values of emission-line EWs lower that that expected for a single burst young population. We will take this into  account while we carry out a detailed comparison of the observed EWs with that predicted from population synthesis models in Sec.4.

\begin{figure}
\begin{centering}
\includegraphics[width=1.0\linewidth]{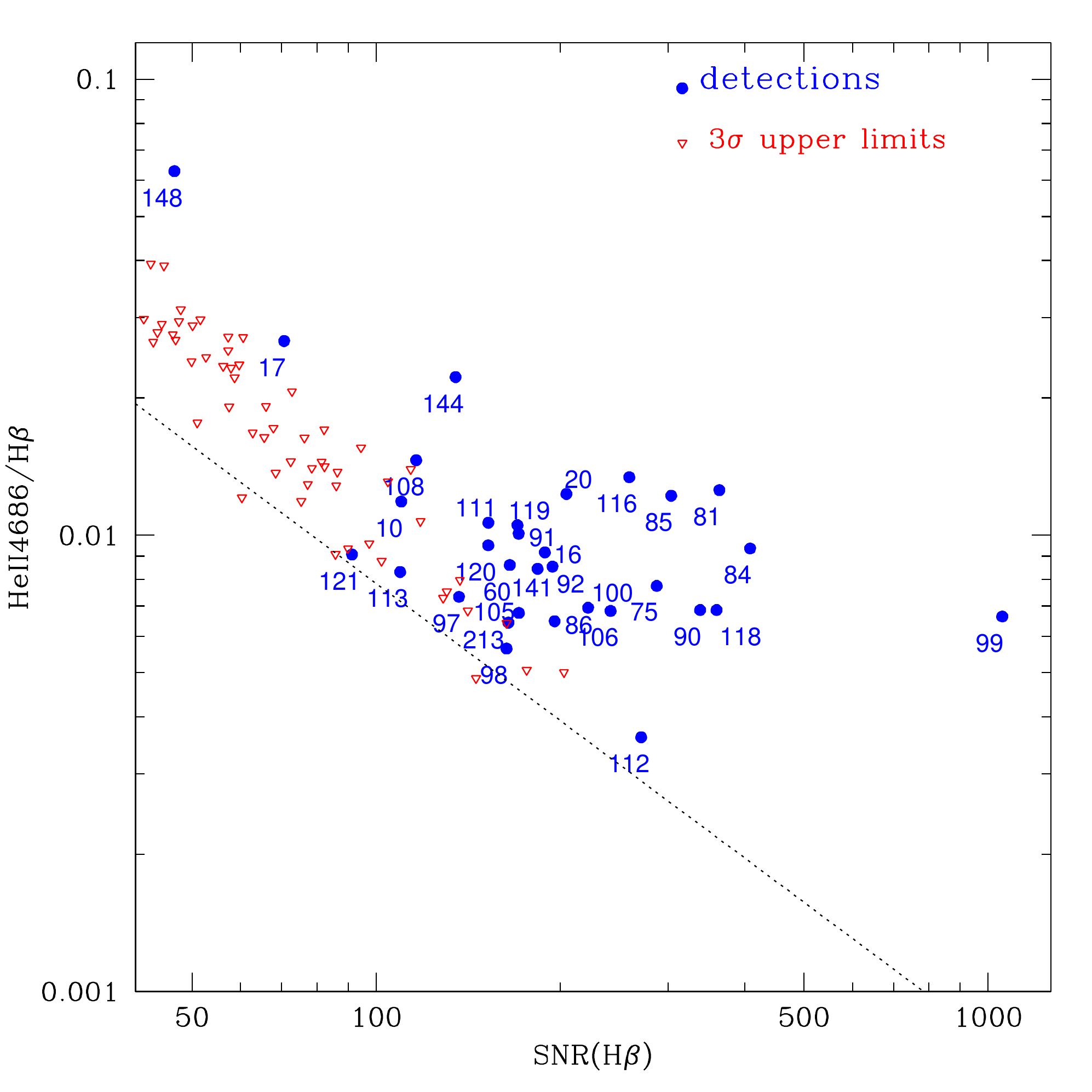}
\par\end{centering}
\caption{The \heiibyhb\ ratio as a function of SNR(\hb) for the Cartwheel \hii\ regions. Detections and the 3-$\sigma$ upper limits for the \heiiwr\ line are shown by blue circles and red triangles, respectively. All detections are tagged by their identification numbers. 
The inclined dotted line denotes the lowest \heiibyhb\ ratio that can be detected at 3-$\sigma$ for \hii\ regions having \ewhb=125~\AA. See text for details.
}
\label{fig:heii_snr}
\end{figure}

\subsection{The He$^{++}$ nebular sample}

We measured fluxes of all major nebular lines in all the extracted spectra.
We used the Gaussian profile fitting routine of the {\it splot} task in {\sc iraf}\footnote{ IRAF is distributed by the National Optical Astronomy Observatories, which
are operated by the Association of Universities for Research in Astronomy,
Inc., under cooperative agreement with the National Science Foundation.}
for this purpose. The measured line fluxes are free from the underlying continuum.
We fitted each extracted spectrum with a Gaussian profile at the redshifted 
wavelengths of a list of nebular lines commonly reported in extragalactic \hii\ regions, 
including the \heiiwr\ line. The output quantities of the Gaussian fitting task that we used here
are: line flux, FWHM and EW. We measured the root mean square (rms) noise at feature-less parts of the continuum adjacent to the measured line, which we used to determine the signal-to-noise ratio (SNR) of the measured line. An emission line is deemed detected if the fitted FWHM is
comparable to that of the \hb\ line (2.5~\AA) and that the integrated line flux has
a SNR$\ge$3. Using these line fluxes, we constructed two sub-samples: (1)
a sub-sample of 87 bright \hii\ regions defined as those with SNR$>$40 in the \hb\ line, and
(2) a sub-sample of 32 \heii\ nebulae defined as those where the \heiiwr\ nebular line is detected.
As expected, the latter sample is a subset of the former. 
The bright \hii\ region sample is used as a control sample to understand
the physical conditions that favour the detection of the \heiiwr\ line in ionized regions.

The criterion that the FWHM of a line should be comparable to that of the \hb\ emission line ensures that the detected line is of nebular origin. Nevertheless, the spectra of the \heii\ nebulae were visually inspected for the possible presence of an underlying broad \heiiwr\ emission feature,
commonly referred to as the 'blue bump' (BB), from WR stars. None of the spectra showed this feature, and hence any contribution of the WR bump to the measured nebular \heiiwr\ fluxes can be ignored.

The \heii\ nebular sample\footnote{Fluxes of all measured lines for the entire sample of \hii\ regions are presented in 
companion paper by \citet{Zaragoza2022}, which deals with the nebular elemental abundances. The ratios of fluxes of lines used in this work are given in a Supplementary Electronic Table.}  is listed in Table~\ref{tab:heii_regions}.
In column 2, we list the cross identification of our sources
with the \hii\ complexes of \citet{Higdon1995}. In other columns,
we give the basic measured quantities of these regions such as the flux, SNR and EW of the \hb\ line 
(columns 4, 5 and 6, respectively), and the \heiiwr\ intensity normalized to I(\hb) and multiplied by 100, along with error (column~9).
The error in measured flux ($\sigma_{\rm l}$) of each line is calculated using
the expression \citep{Tresse1999}:,
\begin{equation}
\sigma_{\rm l} = \sigma_{\rm c} D \sqrt{(2 N_{\rm pix} + \frac{EW}{D})},
\label{eqn:noise}
\end{equation}
where $D$ is the spectral dispersion in \AA\ per pixel (1.25 for MUSE), $\sigma_{\rm c}$ is
the mean standard deviation per pixel of the continuum, which is measured in a
line-free part of the continuum adjacent to each line, $N_{\rm pix}$ is the
number of pixels covered by the line, which is equated to the FWHM of the fitted Gaussian profile.
In all these 32 regions, the \hb\ flux has SNR$>$50, with a
median SNR of 176. The \heiiwr\ line is detected at SNR=3--15.

Detection of a line depends on the SNR of the spectrum. In Figure~\ref{fig:heii_snr}, we show the \heiibyhb\ ratio for all the \hii\ regions with SNR(\hb)$\geq$40. 
Detections are shown in blue solid circles, whereas the upper limits, defined as 3-$\sigma$ fluxes where $\sigma$ is calculated using the error equation~\ref{eqn:noise} with $N_{\rm pix}$=FWHM(\hb), are shown by red inverted triangles. As a reference, we show by the dotted line the minimum value of \heiibyhb\ ratio that a region should have so that the \heiiwr\ line is detected, given the SNR of \hb\ in its spectra. The intercept of this line depends slightly on \ewhb, with the plotted line corresponding to \ewhb=125~\AA. The line would shift upwards (downwards) by $\sim$0.005 for regions with EWs a factor of two lower (higher). 
Our sample of \heii\ nebulae has a mean value of I(\heiiwr)/I(\hb)=0.010$\pm$0.003, 
with the lowest and highest ratios being 0.004 (\#112) and 0.07 (\#148). 
We can expect \heii\ line detection only in \hii\ regions with SNR(\hb)$>$100, if their \heiibyhb\ ratio is close or higher than the mean value of the sample. Conversely, the detection of \heiiwr\ line would require abnormally high ratio of \heiibyhb\  in regions with SNR(\hb)$<$100, such as the case in \#148 and \#17. As expected, the \heiiwr\ line is detected in all MUSE spectra with SNR(\hb)$\geq$200 (12 regions). For regions having 100$<$SNR(\hb)$<$200, the \heiiwr\ line is detected in $\sim$59~per cent (17 of 29) of the \hii\ regions.  

\begin{figure}
\begin{center}
\includegraphics[width=1.0\linewidth]{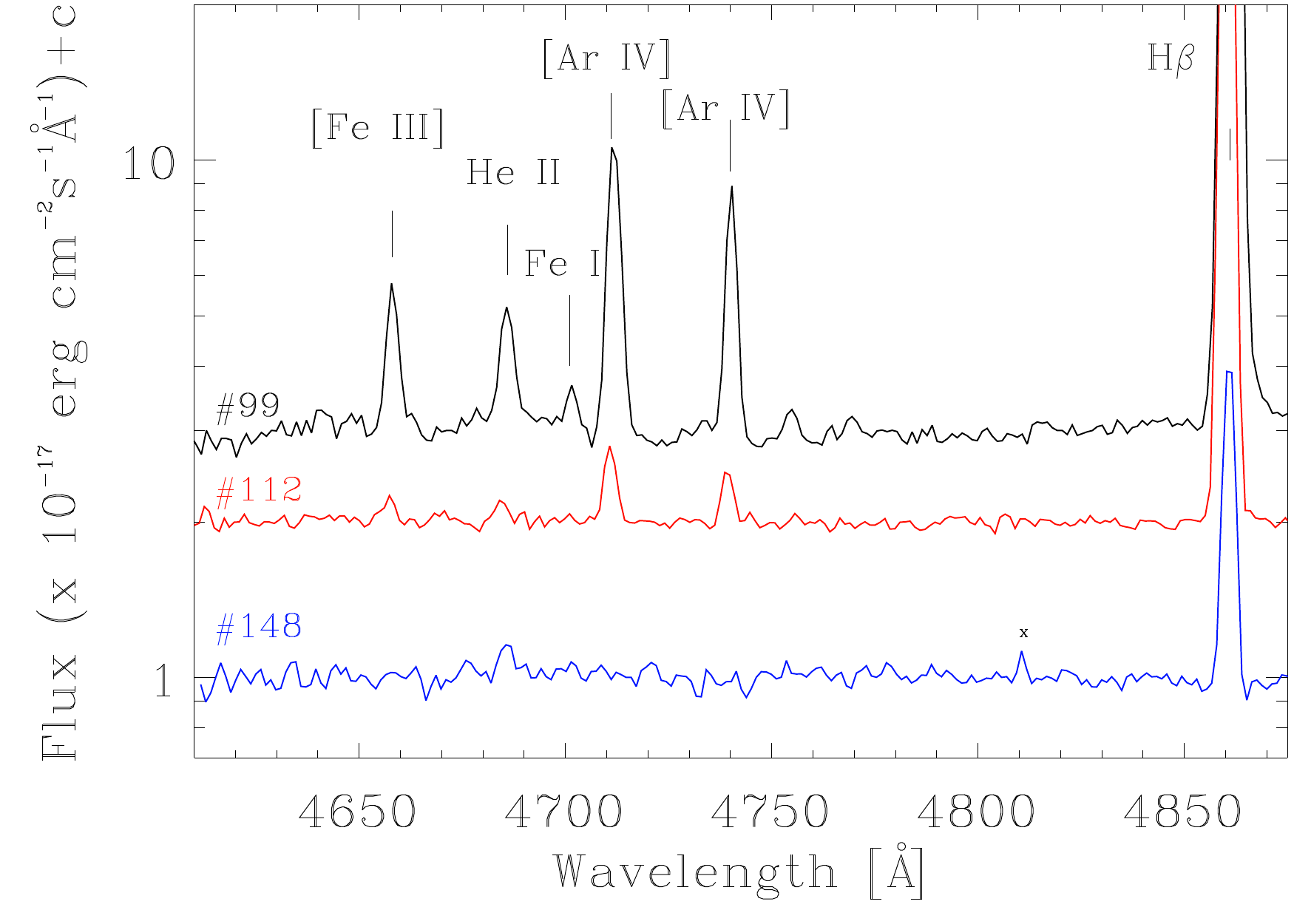}
\caption{The VLT/MUSE rest-frame spectra in the blue part of three representative \hii\ regions \#99, \#112 and \#148 in the Cartwheel. Each spectrum is shown normalized to the best fit continuum spectrum and displaced vertically for clarity sake. The nebular lines in the plotted spectral range are indicated. In the three spectra, the \heiiwr\ line is detected at SNR=12, 3.3, and 3.6 (from top to bottom), with the bottom two spectra illustrating detections just above our chosen 3-$\sigma$ detection limit. In addition, all detections have line widths comparable to that of the typical line spread function. We mark by a cross a spurious feature at $\lambda$=4810~\AA\ in the bottom-most spectrum, which is narrower than the typical line spread function.}
\label{fig:MUSE_spe}
\end{center}
\end{figure}

In Figure~\ref{fig:MUSE_spe} we show the blue part of the MUSE spectrum for three representative \hii\ regions where the \heiiwr\ line is detected. The regions selected for illustrations are: \#99, the brightest \hii\ region in the Cartwheel, \#112 and \#148, the \hii\ regions with the lowest and the highest \heiibyhb\ values, respectively. The \heiiwr\ and other nebular lines in the plotted range are indicated.

\begin{figure*}
\begin{centering}
\includegraphics[width=1\linewidth]{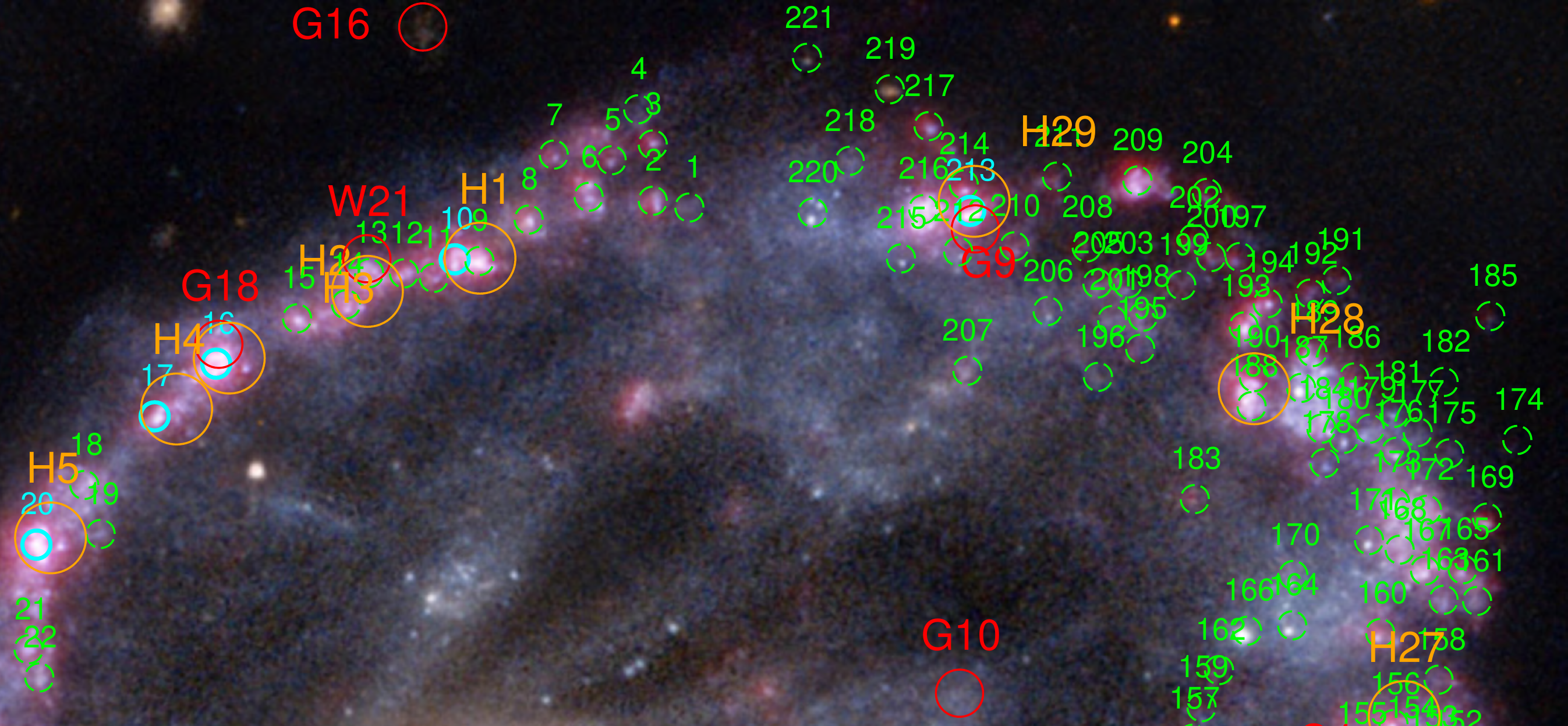}\\
\includegraphics[width=0.244\linewidth]{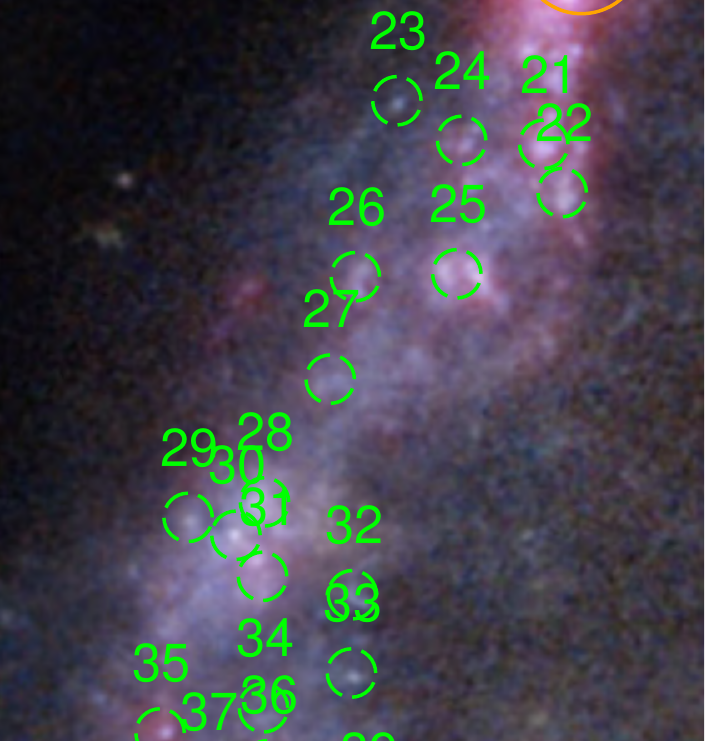}~
\includegraphics[width=0.243\linewidth]{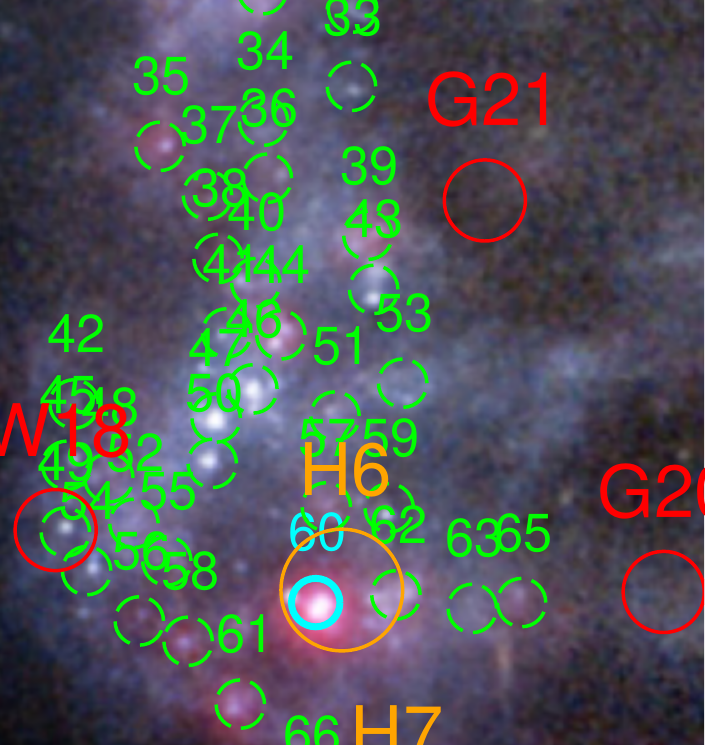}~
\includegraphics[width=0.24\linewidth]{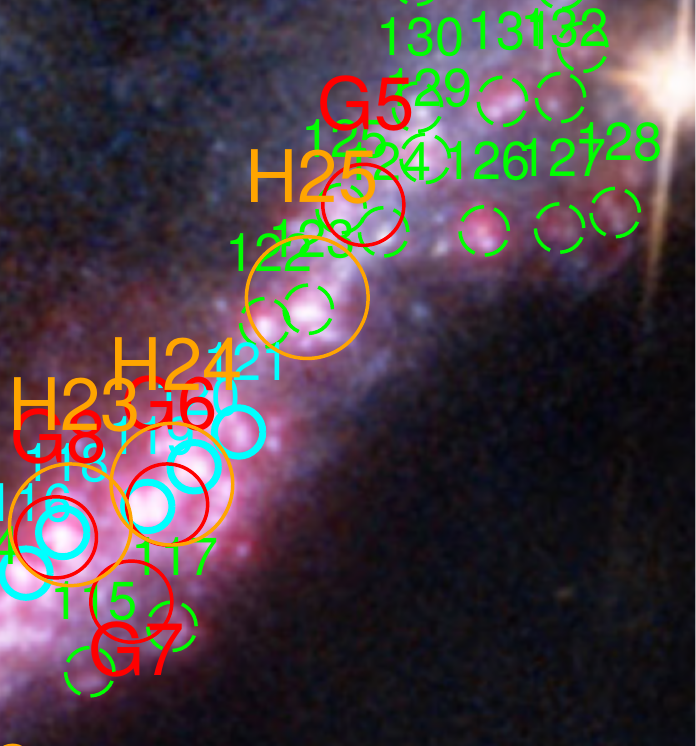}~
\includegraphics[width=0.242\linewidth]{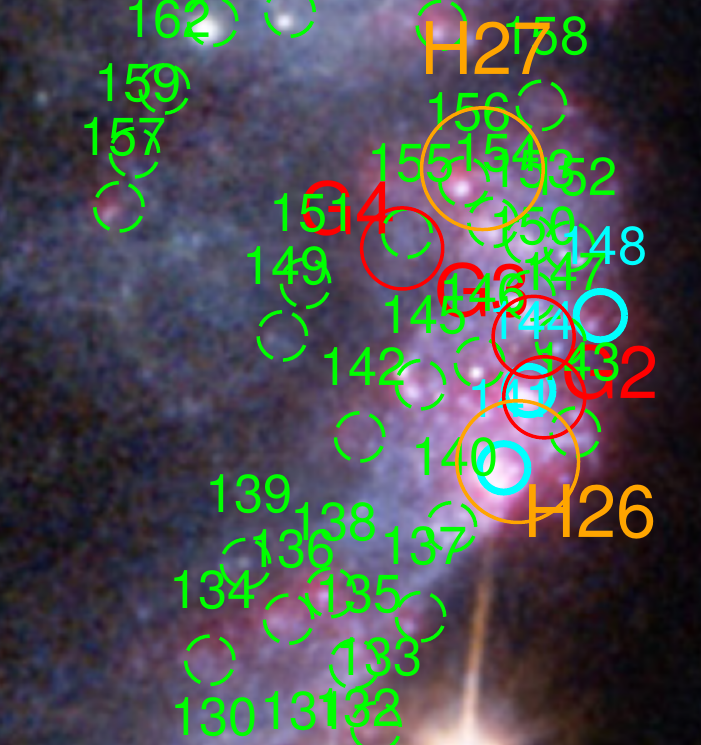}\\
\includegraphics[width=1\linewidth]{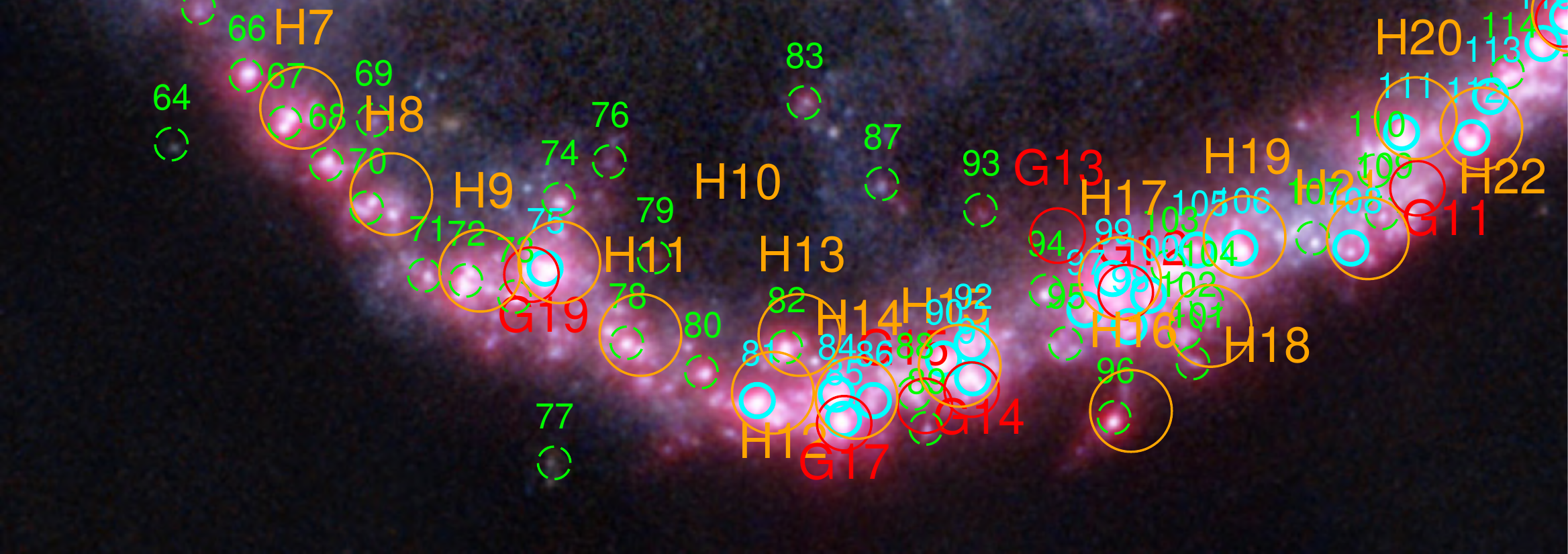}\\
\end{centering}
\caption{Zoomed view of Figure~\ref{fig:muse_image}
to facilitate visualization of the ionizing clusters associated with the MUSE-detected \hii\ regions (small circles) at the HST resolution. Top and bottom images correspond to northern and southern part of the ring, respectively, whereas the middle row contains 2 subimages each of the eastern and western ring.
Thirty two \hii\ regions with well-detected \heiiwr\ nebular line are distinguished
by cyan coloured circles, with the rest of the \hii\ regions identified by
dashed green circles. ULX sources are identified by red circles. The 29 \hii\ complexes defined by \citet{Higdon1995} are shown by big orange circles identified by labels H1 to H29.
The orientation of all sub-images are the same as in Figure~\ref{fig:muse_image},
and small circles in all images have a radius of 0.6~arcsec.
In general, \heiiwr-emitting regions are among the brightest \hii\ regions.
However, there are fainter \hii\ regions (e.g.~\#10, \#148) with \heiiwr\ emission.
See text for details.
}
\label{fig:zoom_image}
\end{figure*}

\subsection{Association of He$^{++}$  nebulae to Star Clusters}

We use the astrometrized HST images in the F450W and F814W filters to look for a spatial association between the He$^{++}$ nebulae and star clusters. In Figure~\ref{fig:zoom_image}, we present a close-up view of the entire ring with the intention of showing the cluster(s) associated to He$^{++}$ nebulae. The HST images reveal the presence of at least one ionizing cluster inside 
the aperture used to extract the MUSE spectrum in all the 32 He$^{++}$ nebulae.
This association suggests that photoionization from stars in young clusters,
or alternatively any process related to young clusters, is mainly responsible 
for the ionization of He$^+$.

\begin{figure}
\begin{centering}
\includegraphics[width=1\linewidth]{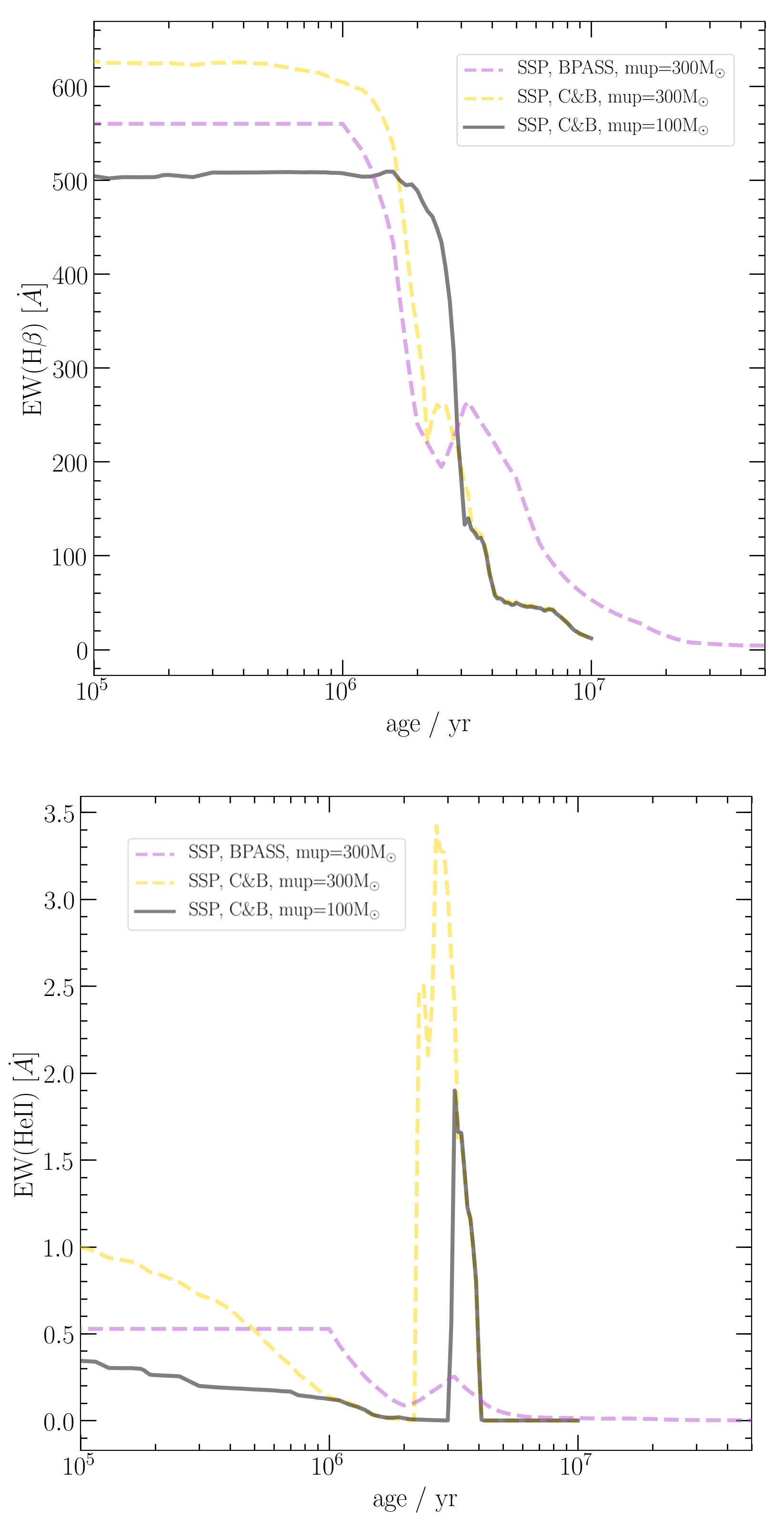}
\end{centering}
\caption{
Evolutionary behaviour of the emission EWs of \hb\ (top) and the nebular  \heiiwr\ (bottom) for C\&B SSPs (single-stars) with $M_{\rm up}$=100~\msol\ (black solid line) and 300~\msol\ (yellow dashed line) and BPASS (binary-stars). See text for details. 
}
\label{fig:EWHb_evol1}
\end{figure}

\begin{figure}
\begin{centering}
\includegraphics[width=1\linewidth]{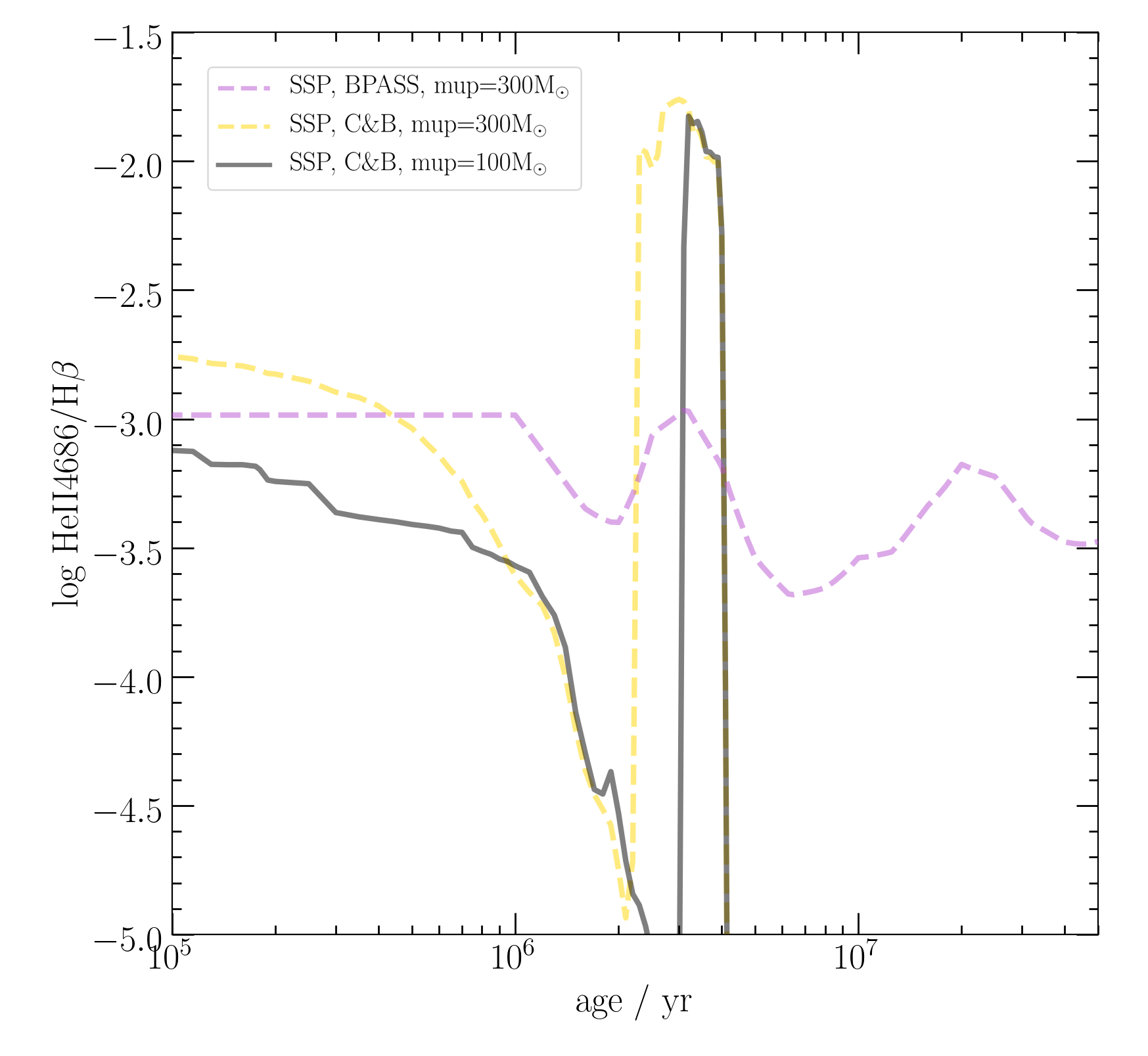}
\end{centering}
\caption{
Evolutionary behaviour of nebular \heiibyhb\ ratio for C\&B (single-star) and BPASS (binary-star) SSPs. See text for details. 
}
\label{fig:EWHb_evol2}
\end{figure}

\begin{figure}
\begin{centering}
\includegraphics[width=1\linewidth]{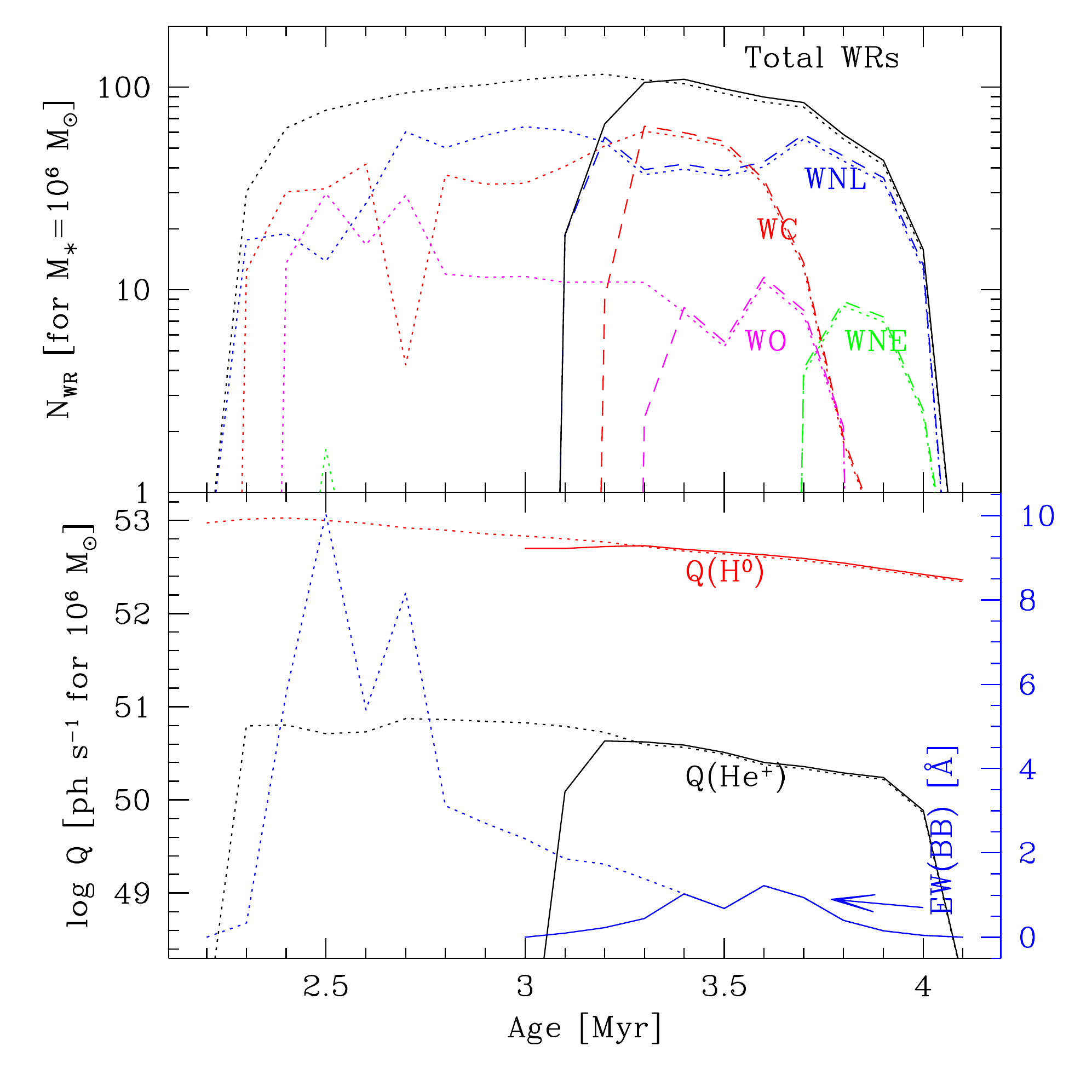}
\end{centering}
\caption{
Evolution of cluster quantities during the WR phase in the C\&B SSP model at Z=0.004. 
The top panel shows the number of different types of WR stars, while the bottom panel shows the ionizing photon rates of H$^0$ and He$^+$ (left axis) and the EW of the Blue Bump (solid blue line and the right axis). 
In both the plots, solid lines correspond to the default model of $M_{\rm u}$=100~\msol\ used in this work, whereas the dotted lines that extend to ages $<$3~Myr correspond to $M_{\rm u}$=300~\msol. In the top panel, the number of different WR subtypes are shown by dashed and dotted lines for $M_{\rm u}$=100 and 300~\msol, respectively.
}
\label{fig:Nwr_CB_evol}
\end{figure}

The star forming complexes defined by \citet{Higdon1995}, after correcting for offsets in their coordinates with respect to our MUSE \ha\ image, are also marked in Figure~\ref{fig:zoom_image}.  The radius of the circles used to identify the \hii\ regions and \hii\ complexes indicates the FWHM of the point spread function (PSF) of the corresponding observations. With the exception of one source (\#148), all are associated with these complexes. In general, these complexes contain multiple \hii\ regions and stellar clusters. However, only 18 of the complexes are associated with a He$^{++}$ nebula, with a few of the complexes containing multiple knots of He$^{++}$ emission. H17 in the southern quadrant of the ring, the brightest complex, contains as much as four knots of He$^{++}$ emission.

\subsection{The \heiiwr\ nebular line in the SSP models}

Photoionization from massive stars is the dominant source of ionization in \hii\ regions. Massive stars with zero-age main-sequence mass $M_\mathrm{ZAMS} \gtrsim 25$~M$_{\odot}$ go through the WR phase during their post-main sequence evolution \citep[see][and references therein]{2007Crowther}. WR stars are the hottest stars in coeval young clusters whose surface temperature reaches in excess of 10$^5$~K, hot enough to doubly ionize helium. The exact age and duration at which WR stars appear in a coeval population depends on the metallicity, the upper cut-off mass ($M_{\rm u}$) of the Initial Mass Function (IMF), and also the inclusion or not of binary star populations in the SSP models used. In this work, we used the latest version of \citet{Bruzual2003} SSP models (Charlot \& Bruzual in-preparation, hereafter C\&B; see also \citet{Plat2019} for details) for single stars corresponding to Z=0.004, the metallicity in the SSP models closest to that of the Cartwheel.
These updated models have incorporated the theoretical spectra from Potsdam Wolf-Rayet (PoWR) model library \citep[see references in][]{Todt2015} and the stellar evolutionary tracks from \citet{Chen2015} that were computed with the {\sc parsec} code of \citet{Bressan2012}. Tracks for Very Massive Stars up to initial masses of 600~\msol\ are available in this code. We here use the SSP models adopting a \citet{Chabrier2003} IMF with lower cut-off mass of $M_{\rm l}$=0.1\,\msol. We present the model results for two values of upper cut-off mass: $M_{\rm u}$=100\,\msol\ and $M_{\rm u}$=300\,\msol. We investigate the effect of including binary star processes (such as envelope stripping and chemical homogenisation) using the BPASS v2.2.1 stellar population models \citep{Stanway2018}.
The models are computed for a zero-age volume-average ionization parameter \logU\ =$-2$ (see Section~4 for details).

In Figures~\ref{fig:EWHb_evol1} and \ref{fig:EWHb_evol2}, we show the evolutionary behaviour of the EWs for the \hb\ and \heiiwr\ nebular lines, and the intensity ratio, I(\heiiwr)/I(\hb). For C\&B models, we show the results for two values of $M_{\rm u}$, 100 and 300\,\msol, whereas the binary models are shown only for $M_{\rm u}$=300\,\msol.
The nebular \ewhb\ steadily decreases with age as massive stars die in a coeval population. On the other hand, the nebular  EW(\heiiwr) shows a second peak after the initial steady decrease, with the value of this second peak, which corresponds to the WR phase, higher than the values at the pre-WR phase for the C\&B models.
The log(I(\heiiwr)/I(\hb)) has values between $-$3.5 to $-$2.7 when all stars are in the main sequence. However, the highest value is reached during the WR phase of single star evolutionary models, which happens between 2.3 to 4.0~Myr with $M_{\rm u}$=300~\msol, and between 3.2 to 4.0~Myr with $M_{\rm u}$=100~\msol. In C\&B models, the highest value is close to $-1.8$ for both the values of $M_{\rm u}$, dropping steeply to values below $-5$ at 4~Myr.
The evolutionary trend is notably different in BPASS binary models: the highest values are reached during the main sequence phase, thereafter the value remaining between $-$3.5 and $-$3 to ages even beyond 30~Myr.
Unfortunately, the highest value of I(\heiiwr)/I(\hb) ratio differs in different publicly available SSP models, with the value depending on the details of the stellar evolutionary and atmospheric models used as illustrated in \citet{Mayya2020}. The mean value of I(\heiiwr)/I(\hb)=0.0094$\pm$0.0025 for our  sample regions is in good agreement with the expected value for ionization from WR stars in C\&B models, whereas it is higher than the peak value reached in BPASS models. The systematically low value of this ratio during the WR phase in BPASS single star and binary star models limits the use of BPASS models, as of now, for a robust discussion of WR stars as ionizing source in our \heii\ nebula. We hence draw conclusions regarding the source of ionization of He$^+$ using the C\&B SSP models, and make use of the BPASS models only to discuss qualitative changes in our C\&B-based results, if the Cartwheel clusters contain binaries in significant numbers. 

In Figure~\ref{fig:Nwr_CB_evol}, we show the expected number of WR stars of different sub-types during the WR phase (top panel), the ionizing photons rates Q(H$^0$) and Q(He$^+$), as well as the expected EW of the BB (bottom panel) for the Z=0.004 C\&B SSP. 
Note that these numbers do not take into account the presence of WR-like stripped (or spun-up) binary stars, which are capable of ionizing He$^+$, but do not show strong blue bumps \citep[see e.g.][]{Gotberg2018}. 
In this figure, all quantities except EW(BB) depend on the cluster mass. We scaled all calculated values to a cluster mass of $10^6$~\msol. In order to calculate the model EW(BB), we obtained the flux in the BB, and the underlying continuum. The bump flux is obtained by integrating the model spectra of pure WR stars between 4570\,\AA\ and 4740\,\AA. The underlying continuum flux is obtained as a mean of continuum fluxes at the blue and red part of the BB. The blue and red continua are obtained over 50\,\AA\ width filters centered at 4525\,\AA\ and 4905\,\AA, respectively.

Both the $M_{\rm u}$=100~\msol\ and 300~\msol\ models have a total number of $\sim$100~WR stars/$10^6$~\msol\ of stellar mass at any time during the WR phase. The WC and WNL stars are the two main types expected up to around 3.6~Myr, after which WNL type is the dominant type with some contribution from WNE type. However, as illustrated earlier, the WR phase begins earlier for higher $M_{\rm u}$, which leads to earlier availability of the He$^+$ ionizing photons for $M_{\rm u}$=300~\msol\ models.  
It can be noticed from the figure that the Q(He$^+$) increases by more than two orders of magnitudes when the WR stars start appearing in the cluster, whereas the Q(H$^0$) continues its steady decrease that started at the end of the pre-WR phase into the WR phase. 
Unlike Q(He$^+$) and Q(H$^0$), whose maximum values do not depend on the chosen $M_{\rm u}$ of the IMF, the maximum value of the EW(BB) is very sensitive to the choice of $M_{\rm u}$. The maximum value reached for $M_{\rm u}$=100~\msol\ is 1.2~\AA\ at 3.6~Myr, whereas it is as high as 10~\AA\ for the $M_{\rm u}$=300~\msol\ models. The evolutionary behaviours for $M_{\rm u}$=300~\msol\ and 100~\msol\ are indistinguishable for ages greater than 3.4~Myr.

\subsection{Ionizing photon rates and cluster masses}

We used the observed luminosity in the \hb\ recombination line, L(\hb), to estimate the equivalent number of O7V stars using a typical luminosity of an O7V star of 4.76$\times10^{36}$~\ergs\ from \citet{Esteban2010}.
We also calculated the H$^0$ and He$^+$ ionizing photon rates, Q(H$^0$) and Q(He$^+$), respectively, for a Case B photoionized nebula using the basic equations for photoionized nebulae \citep{Osterbrock2006, Mayya2020}. 
We then calculated the stellar mass in the cluster that is ionizing each region, using:\\
\begin{equation}\label{eqn:QHe}
\frac{M_\ast}{\rm M_\odot} = \frac{{\rm Q(H}^0)}{{{\rm Q(H}^0)_{\rm SSP}}},
\end{equation}
where ${{\rm Q(H}^0)_{\rm SSP}}=4.94\times10^{46}~{\rm photon~s^{-1}}$\msol$^{-1}$ is the H$^0$ ionizing photon rate for the Z=0.004 SSP in the C\&B models for the characteristic age in the WR phase of 3.4~Myr. The masses would be around a factor of two lower, or higher, for the clusters that are in the pre- or post-WR phases, respectively.  
The derived values of N(O7V), M$_\ast$, and Q(He$^+$) are given in columns 10, 7 and 8, respectively, of Table~\ref{tab:heii_regions}.

\subsection{Potential ionizing sources of He$^{++}$ nebulae}

\subsubsection{Wolf-Rayet stars}\label{sec:WRmodel}

From Figure~\ref{fig:EWHb_evol2}, it is apparent that the WR stars are the most likely sources of ionization of He$^+$ in majority of our \hii\ regions. However, as commented earlier in this section, none of the spectra showed the BB (neither other WR features; see e.g. \citealt{Gomez-Gonzalez2021}). In order to understand this apparent contradiction, we estimated an upper limit on the EW of the BB,  using the noise measurement in the continuum adjacent to the expected BB and assuming a Gaussian profile of FWHM=20~\AA\ for the BB \citep[see e.g.][]{Gomez-Gonzalez2020}. The resulting EWs are given in column~11 of Table~\ref{tab:heii_regions}. In the table, we also give the theoretically expected number of WNL-type and all WR stars in each region, assuming a uniform age of 3.4~Myr. The expected number of WR stars is as high as 242 for region\#99, and in the 30--40 range for four other regions (\#90, 81, 84, 118). In rest of the regions the expected numbers are less than 30.

\begin{figure}
\begin{centering}
\includegraphics[width=1\linewidth]{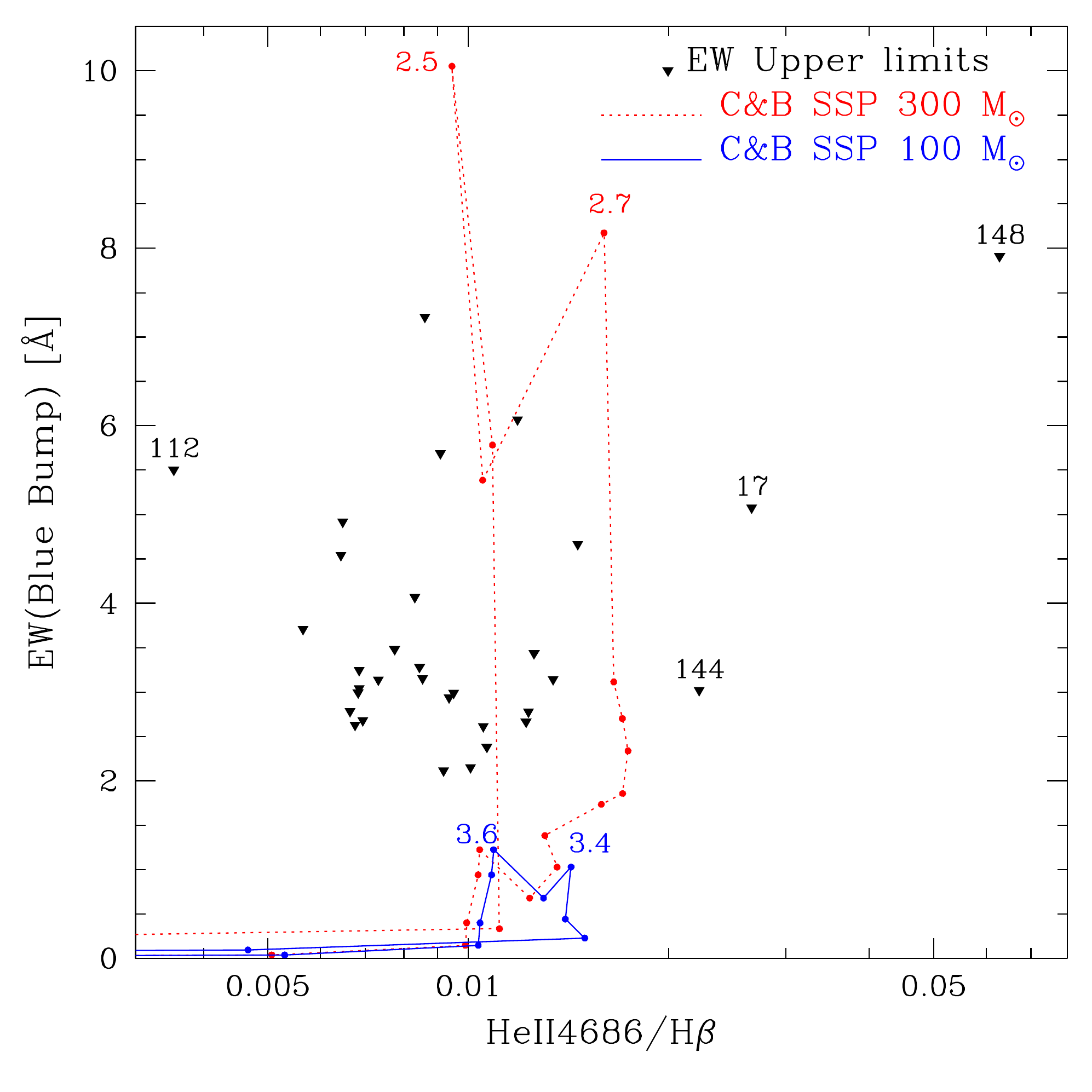}
\end{centering}
\caption{
Observational upper limits (inverted triangles) on the EW(BB) plotted against the \heiibyhb\ ratio for Cartwheel \hii\ regions with \heiiwr\ line detections. The expected values during the WR phase from the C\&B SSP model for two upper cut-off masses are shown. In the presence of massive stars of 300~\msol\ the EW(BB) reaches values as high as 10~\AA\ for a brief period around 2.5~Myr, whereas the EW(BB) stays lower than 1.5~\AA\ when $M_{\rm u}$=100~\msol, which is at least a factor of two lower than the detection threshold as indicated by the 3-$\sigma$ upper limits. \hii\ regions with extreme values of \heiibyhb\ ratio are identified with their numbers. See text for details.
}
\label{fig:Nwr_vs_QHeII}
\end{figure}

The observed upper limits on the EW(BB) are compared with those expected in the C\&B models in the EW(BB) vs \heiibyhb\  plane in Figure~\ref{fig:Nwr_vs_QHeII}.
As discussed earlier, the theoretically expected EW(BB) reaches a maximum value of only 1.2~\AA\ for $M_{\rm u}=100$~\msol, whereas it could be as high as 10~\AA\ for $M_{\rm u}=300$~\msol. 
The observational detection upper limits on the EW(BB) are higher by factors of 2 to 6 as compared to the theoretical values in the absence of Very Massive  (mass$>$100~\msol) Stars. This illustrates that the absence of the blue bump does not necessarily imply the absence of WR stars. At the metallicity of the Cartwheel \hii\ regions, WR stars in numbers sufficient to doubly ionize helium could be present even when the characteristic broad BB is not detected. This could be the case not only in the Cartwheel regions, but in general in all metal-poor systems that show \heiiwr\ nebular line \citep[e.g.][]{Shirazi2012}.

It is interesting to note that we would have been able to detect the BB when Very Massive Stars in clusters, if present, go through the WR phase. 
It can be inferred from Figure~\ref{fig:Nwr_CB_evol}, that the $M_{\rm u}$=300~\msol\  models provide He$^+$ ionizing photons for double the duration as compared to the $M_{\rm u}$=100~\msol\ models. This implies that if all the \heiiwr-detected \hii\ regions had an IMF with $M_{\rm u}$=300~\msol, the BB would have been at the detectable level in 50\% of the cases. The non-detection of the BB in all regions points to the absence of $M_{\rm u}>$100~\msol\ stars in \hii\ regions of the Cartwheel.

Given that our \heiiwr-line detection criteria are tuned to detect nebular (narrow) lines, there exists a possibility that we inadvertently excluded possible WR sources in spectra where we did not detect the \heiiwr\ narrow line. In order to verify this possibility, we analyzed the output results of the Gaussian-fitting for all the regions to look for a broad \heiiwr\ component with SNR$\ge$3. None of the spectra showed evidence for it. If stars more massive than 100~\msol\ were common, the BB should have been present at detectable levels in at least a few of the 80 \hb-bright regions. In fact, none of the spectra of our original sample of 221 \hii\ regions showed the BB. Such a non-detection reinforces the inference drawn from the \heiiwr-detected regions that the upper mass cut-off of the IMF rarely exceeds 100~\msol\ in \hii\ regions. 

In summary, WR stars are viable sources of He$^+$ ionization in \heiiwr-detected regions of the Cartwheel, in spite of the non-detection of the BB. In rest of this paper, we use photoionization models to investigate whether the intensity ratios of bright nebular lines support the scenario of WR stars as the only source of ionization in majority of the regions. Four regions with extreme \heiibyhb\ ratio (identified in Figure~\ref{fig:Nwr_vs_QHeII}), possibly require alternative sources of ionization, which are also investigated in the paper.

\subsubsection{Main sequence stars}

The bulk of the ionization of hydrogen in \hii\ regions is provided by star clusters in their early phase when massive O stars are in the main sequence (MS). Some of these stars are hot enough to doubly ionize helium. The emission \ewhb\ is maximum during this early phase having values larger than 500~\AA. 
The highest values of \heiibyhb\ during this phase are 0.002 and 0.001, respectively in C\&B and BPASS models, both with $M_{\rm u}$=300~\msol. It can be inferred from Figure~\ref{fig:heii_snr} that we require SNR above 400 to detect \heiiwr\ line ionized by the MS stars. There are 5 regions with SNR$>$330, all of which have \heiibyhb$>$0.006, i.e. at least a factor of three higher than the MS values. On the other hand, the region with the lowest value of \heiibyhb\ is \#112, which is around twice the MS value.
Stripped binary stars, which are not taken into account in the C\&B and BPASS models, may have a role in increasing the \heiibyhb\ above the calculated values. The MS phase is characterized by high \ewhb. We analyse all regions in \heiibyhb\ vs \ewhb\ plane to address this issue in Sec.~3.

\begin{table}
\small\addtolength{\tabcolsep}{-3pt}
\caption{Cartwheel \hii\ regions nearest to an ULX source.}
\label{tab:ulx_regions}
\begin{tabular}{rllcclccl}
\hline
\hii\ ID & Higdon & \multicolumn{2}{c}{Gao} & & \multicolumn{3}{c}{Wolter} & He$^{++}$\\
\cline{3-4}
\cline{6-8}
    &     & ID   & offset&&ID& offset & log L$_{\rm X}$   \\
    &     &      &[arcsec]&& &[arcsec]&  erg\,s$^{-1}$ &  \\
(1) & (2) & (3)  & (4)  && (5) & (6)  & (7) & (8) \\
\hline
 13 & H2  &  --- & ---  && W21 & 0.61 & 38.70 & no  \\
 16 & H3  &  G18 & 0.84 && --- & ---  & --- & yes \\
 49 & --- &  --- &  --- && W18 & 0.29 &       & no  \\
 75 & H10 &  G19 & 0.65 && W23 & 0.33 & 38.78 & yes \\
 85 & H14 &  G17 & 0.16 && W7  & 0.42 & 39.70 & yes \\
 88 & --- &  G15 & 0.12 && W8  & 0.52 & 39.18 & no  \\
 91 & H15 &  G14 & 0.44 && --- & ---  & --- & yes \\
 99 & H17 &  G12 & 0.77 && W24 & 2.39 & 38.57 & yes \\
111 & H20 &  G11 & 2.17 && W10 & 1.75 & 40.44 & yes \\
117 & --- &  G7  & 1.33 && W13 & 1.21 &   & no  \\
118 & H23 &  G8  & 0.24 && --- & ---  & --- & yes \\
119 & H24 &  G6  & 0.56 && W14 & 1.28 & 39.88 & yes \\
125 & --- &  G5  & 0.65 && W15 & 1.42 & 39.06 & no  \\
144 & H26 &  G2  & 0.50 && W17 & 0.70 & 39.92 & yes \\
146 & --- &  G3  & 1.74 && W16 & 2.32 & 39.83 & no  \\
155 & --- &  G4  & 0.41 && --- & ---  & --- & no  \\
213 & H29 &  G9  & 0.81 && W12 & 0.72 & 39.49 & yes \\
\hline
\end{tabular}
\end{table}

\subsubsection{Ultra-luminous X-ray sources}\label{sec:ulx}

Pointlike non-nuclear X-ray sources with isotropic bolometric luminosity in the the 0.5--10~keV band ($L_{\rm X}$) exceeding 3$\times10^{39}$~\ergs are referred to as ULX sources. 
\citet{Gao2003} and \citet{Wolter2004} analyzed the {\sc Chandra/Acis}
data of Cartwheel finding 31 and 24 ULX sources, respectively, in the
FoV of the Cartwheel, the majority of them coinciding with the star-forming
ring of the Cartwheel. The most luminous of these sources (\#11 in \citealt{Gao2003} 
and \#10 in \citealt{Wolter2004}) has $L_{\rm X}>10^{41}$~\ergs, thus satisfying
the criterion to be called as a hyperluminous X-ray source (HLX). 
However, a one-to-one correspondence with an optical knot in the ring was poor 
in both the studies, with offsets between the \hii\ complexes defined by 
\citet{Higdon1995} and the ULX coordinates, generally exceeding the 1~arcsec 
beam of the {\sc Chandra/Acis} observations, even after correcting for zeropoint 
offsets in the respective coordinate systems. The reason for 
these large offsets is that there is more than one star cluster within the seeing-limited resolution of $\sim$1.7~arcsec (1~kpc) of the \ha\ image of \citet{Higdon1995}, with the coordinates referring to that of the brightest \hii\ region in the complex, which is not necessarily the ULX source.
The astrometrically calibrated MUSE and HST dataset that we use in this study offers $\sim$3 and 8 times better spatial resolutions, respectively, as compared to the \ha\ image of \citet{Higdon1995}, which allows us to improve upon the identification of the optical counterpart of the ULX sources.

In Figure~\ref{fig:zoom_image}, we mark the positions of ULX/HLX sources by red circles
overlaid on the HST image. Fourteen of the 17 X-ray sources coincide with
the position of an \hii\ region to better than an arcsec, the beam of
the X-ray observations. The \hii\ region closest to a ULX/HLX source
is identified in Table~\ref{tab:ulx_regions}, where we also give the
offsets for each source from the coordinates reported by \citet{Gao2003} and 
\citet{Wolter2004}. It is worth noting that given the high density of \hii\ regions in
the ring, more than one \hii\ region can be associated for sources
with offsets exceeding an arcsec. The offsets are systematically smaller for \citet{Gao2003}
coordinates. Source \#111, the identified counterpart of the HLX source (G11) is outside the  {\sc Chandra/Acis} beam, suggesting that the identification is likely to be wrong, and that the source may be associated to a non-\ha-emitting object. In order to identify such a candidate, we looked for any stellar knot  in the HST images. We find a faint red knot at the edge of the {\sc Chandra/Acis} beam, which could be a likely counterpart of the HLX source (see the region G11 in the figure).
A He$^{++}$ nebula is present within the beam of the X-ray observations for
ten and seven ULX sources identified by \citet{Gao2003} and \citet{Wolter2004}, respectively (see the last column of the table). We analyse below the possible role of X-rays from the ULX sources in the ionization of He$^+$.

\citet{Schaerer2019} found that the observed \heiibyhb\ ratio in metal-poor galaxies can be explained if the bulk of the He$^+$ ionizing photons is emitted by HMXBs, whose numbers are found to increase with decreasing metallicity. They obtained an empirical relation between Q(He$^+$) and the X-ray luminosity, ${\rm L_X}$, suggesting an almost constant ratio $q={\rm Q(He^+)/L_X}=2\times10^{10}$~photon\,erg$^{-1}$, with extreme values of $q$ being 1--3$\times10^{10}$~photon\,erg$^{-1}$. \citet{Plat2019} warned that this process is not efficient at \ewhb$>$200~\AA, as these systems are too young to form compact objects (neutron stars and stellar mass black holes) necessary for the existence of HMXBs. Only two Cartwheel \heii-emitting regions have \ewhb$>$200~\AA, and hence ionization of He$^+$ by ULX sources is a possibility in majority of the \heii-emitting regions with an associated ULX source.

\begin{figure}
\begin{centering}
\includegraphics[width=1.0\linewidth]{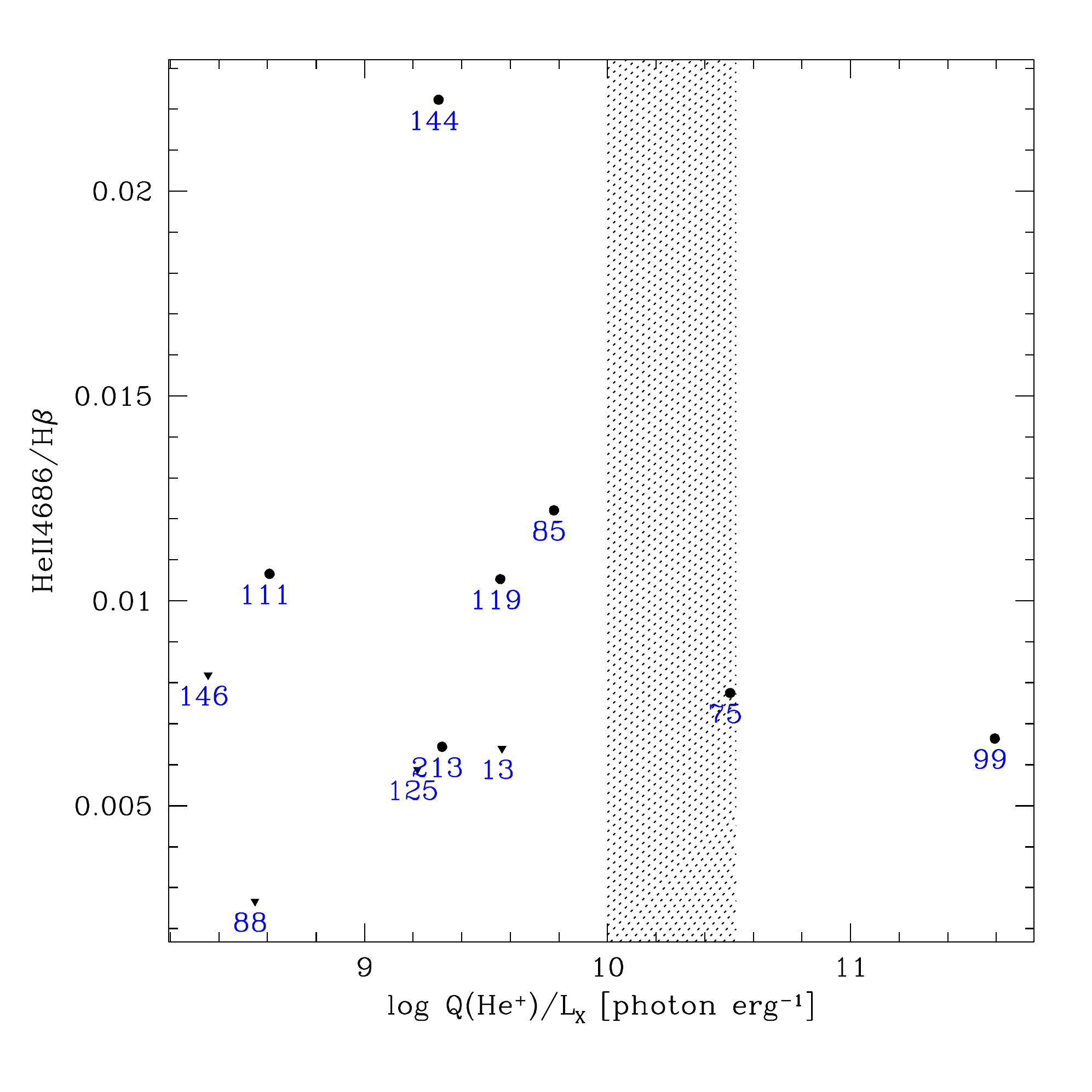}
\par\end{centering}
\caption{
Analysis of the possibility of ionization of He$^+$ by the ULX sources associated with the \hii\ regions  of Cartwheel in \heiibyhb\ ratio vs. $q={\rm Q(He^+)/L_X}$ plane. Circles are detections, and inverted triangles are 3-$\sigma$ upper limits on \heiiwr\ detection. Regions in which The range of values for this ratio proposed by \citet{Schaerer2019} is shown by the vertical hatched area. The points are annotated with the optically identified \hii\ region numbers associated with each ULX source (see Table~\ref{tab:heii_regions}). }
\label{fig:heii_qx}
\end{figure}

\begin{figure*}
\begin{centering}
\includegraphics[width=0.75\linewidth]{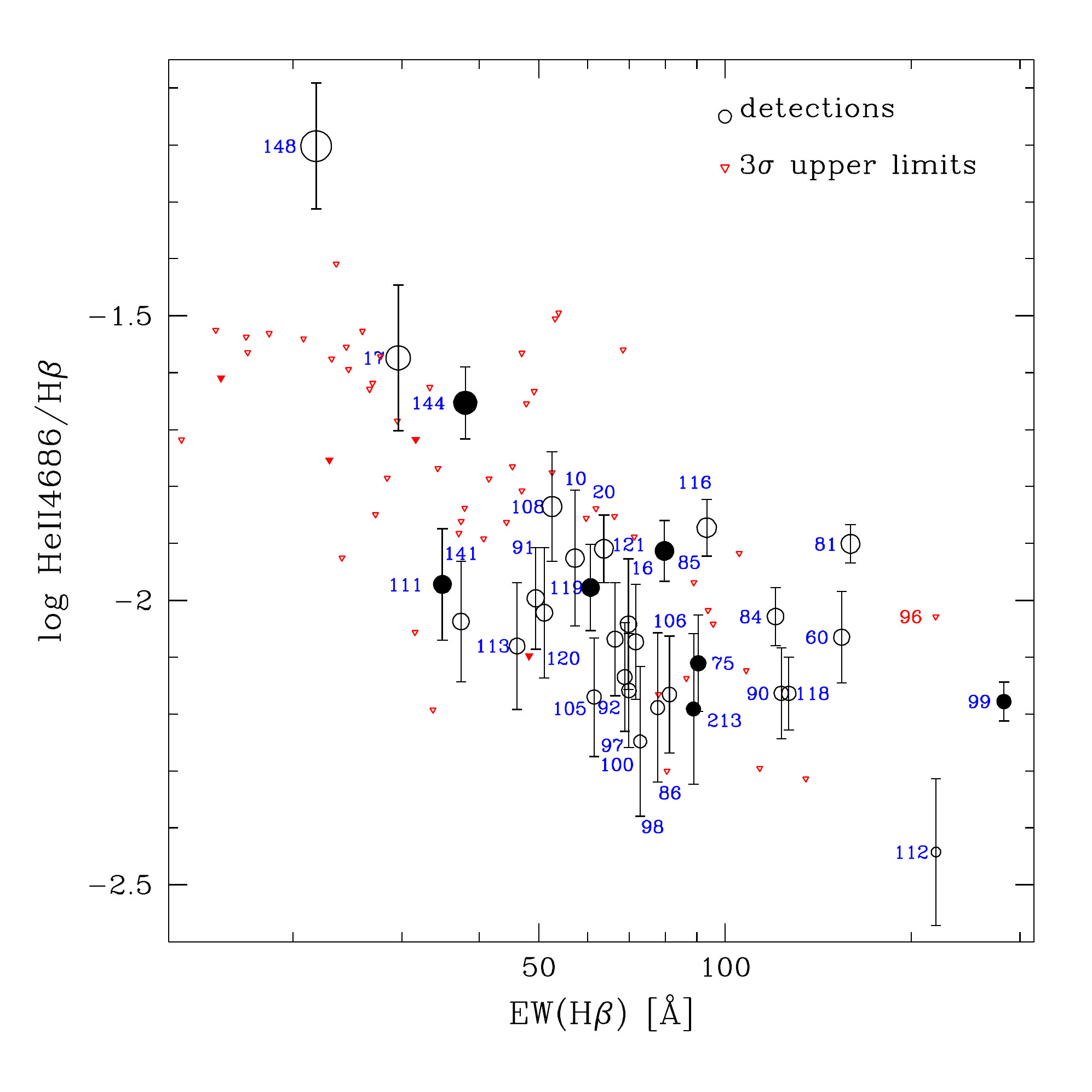}
\par\end{centering}
\caption{Relation between the \heiibyhb\ ratio and \ewhb\ for the Cartwheel \hii\ regions: circles are detections whereas triangles are 3-$\sigma$ upper limits. Filled symbols correspond to \hii\ regions associated to ULX sources. The radius of the circle is proportional to its log(\heiibyhb) ratio.
All \hii\ regions where the \heiiwr\ line is detected as well as those without \heii\ detection, but with 3-$\sigma$ upper limits of \heiibyhb$<$0.06 are identified by their \hii\ region number.
}
\label{fig:heii_ew}
\end{figure*}

The presence of an ULX source in 17 of the Cartwheel \hii\ regions allows us to calculate the value of $q$ directly for these regions. For this purpose, we use the Q(He$^+$) for each region in column~8 of Table~\ref{tab:heii_regions} and the ${\rm L_X}$ of column~7 of Table~\ref{tab:ulx_regions}, which was taken from \citet{Wolter2004}. In Figure~\ref{fig:heii_qx}, we plot the \heiibyhb\ ratio against the $q$ values for the 11 sources for which we have well-determined values of ${\rm L_X}$. The \heiiwr\ line is detected in seven of these sources, with the remaining four only having an upper limit for the detection of the \heiiwr\ line. We find a dispersion of more than 2 orders of magnitudes in the value of $q$ for the individual \hii\ regions in the Cartwheel, with only one of these regions having $q$ in the range found by \citet{Schaerer2019}. This large variation in the $q$ value suggests that the X-rays cannot be the unique source of ionization in these sources. It is likely that not all ULX sources in the Cartwheel are HMXBs, and instead the X-ray luminosity may be originating in supernova (SN) remnants. \citet{Wolter2018} discuss them as HMXBs, whereas in the original detection papers \citep{Gao2003, Wolter2004}, such a possibility was firmly established for the only HLX source (\#111) in the Cartwheel. The $q$-value obtained for this source is more than an order of magnitude lower than the value proposed by \citet{Schaerer2019}. We analyse the fluxes of lines from high ionization levels such as \ariv\ to address the role of X-ray ionization in each of the \hii\ regions associated with an ULX source.

\begin{figure*}
\begin{centering}
\includegraphics[width=0.495\linewidth]{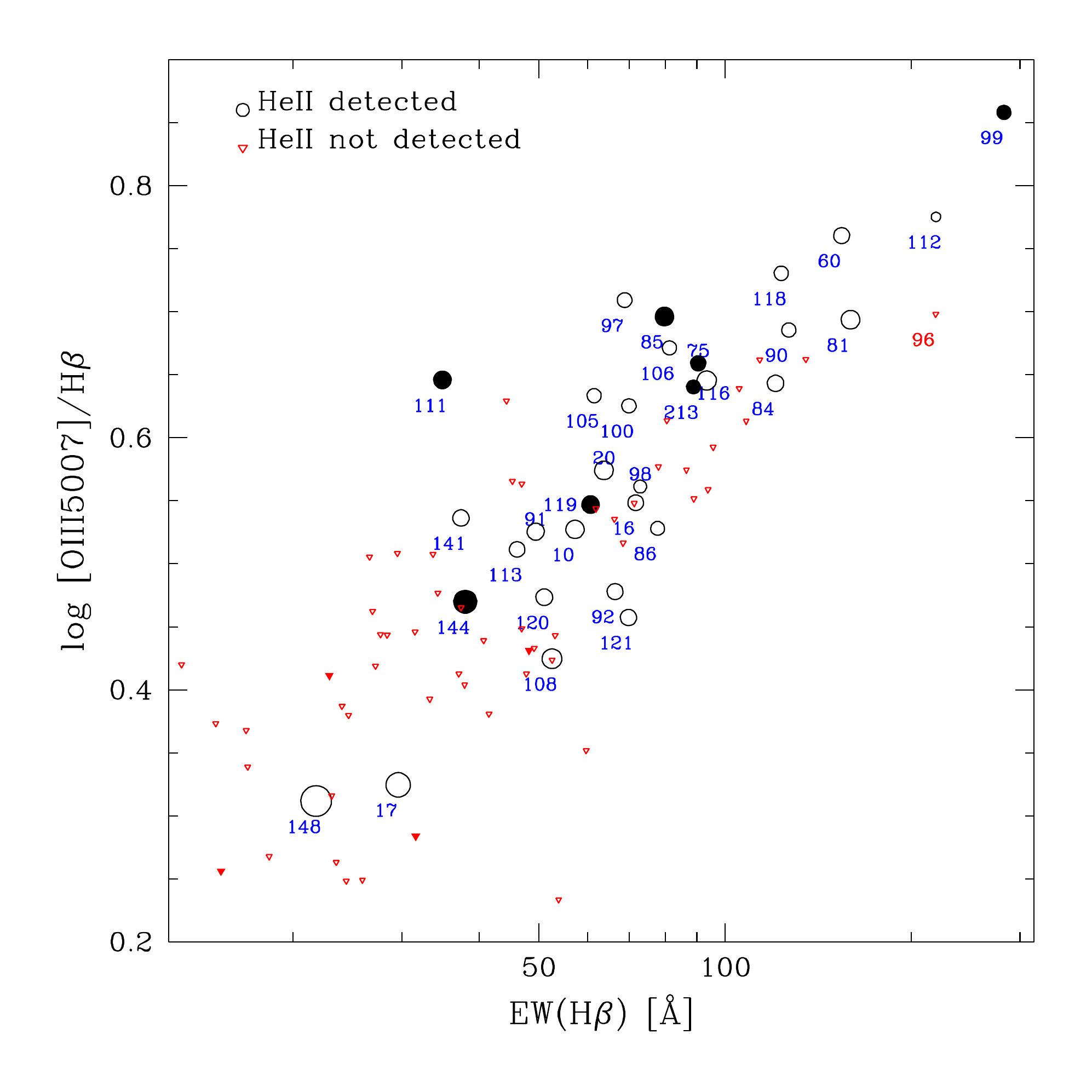}
\includegraphics[width=0.495\linewidth]{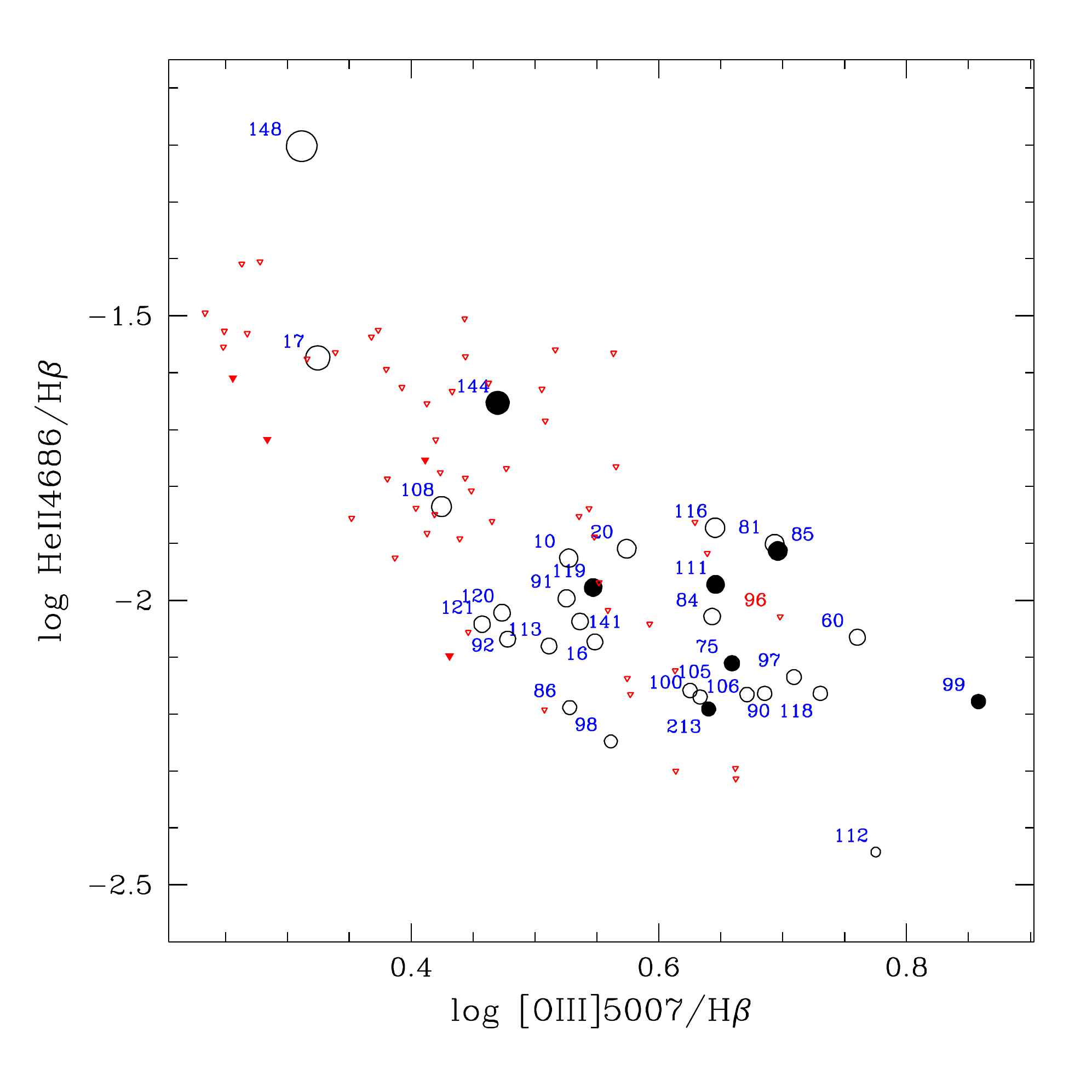}
\par\end{centering}
\caption{(left) \oiiibyhb\ ratio vs. \ewhb; (right) \heiibyhb\ vs. \oiiibyhb\ ratio, plotted against each other. Symbols have the same meaning as in Figure~\ref{fig:heii_ew}.
}
\label{fig:heii_oiii}
\end{figure*}

\section{Analysis of nebular line ratios}

The wealth of spatial and spectroscopic information contained in the MUSE data of the Cartwheel
offers us a great opportunity to comprehensively address the nature of ionizing 
sources based on the ionization state of the nebulae. Specifically, the data
allow us to study whether the \hii\ regions containing the \heiiwr\ line
show any difference with respect to our control sample of 87 \hii\ regions 
in the same galaxy, in any of the diagnostic line ratios commonly used
in ionized nebulae \citep{BPT1981}. 

In this section, we discuss the general trends seen in different line ratio diagrams. We also discuss the possible role of ULX sources in the ionization of He$^+$. In Sec.4, we compare the observed trends with that expected from different theoretical scenarios of ionization of He$^+$.

\subsection{\heiibyhb, \oiiibyhb\ and \ewhb}

\begin{figure*}
\begin{centering}
\includegraphics[width=0.495\linewidth]{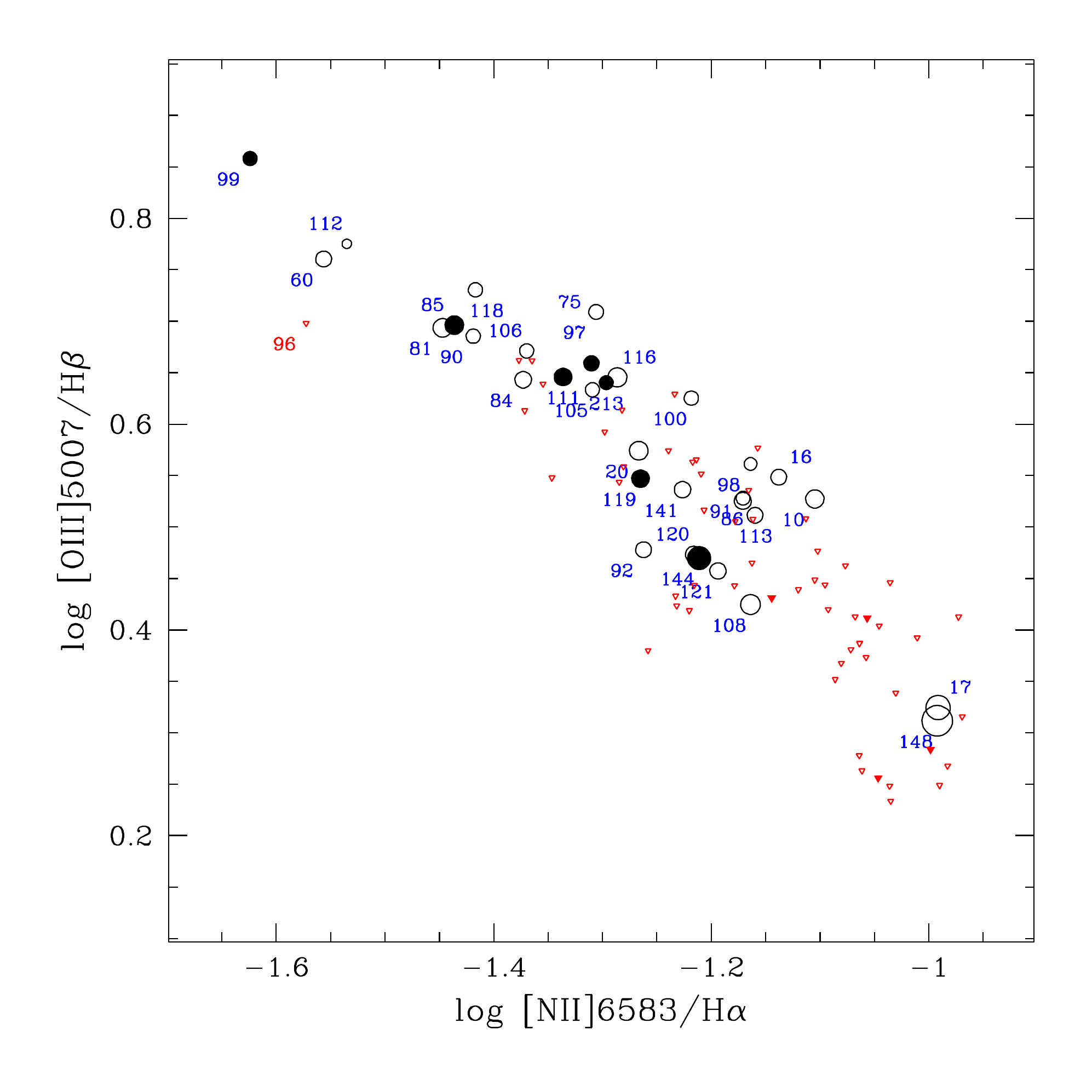}
\includegraphics[width=0.495\linewidth]{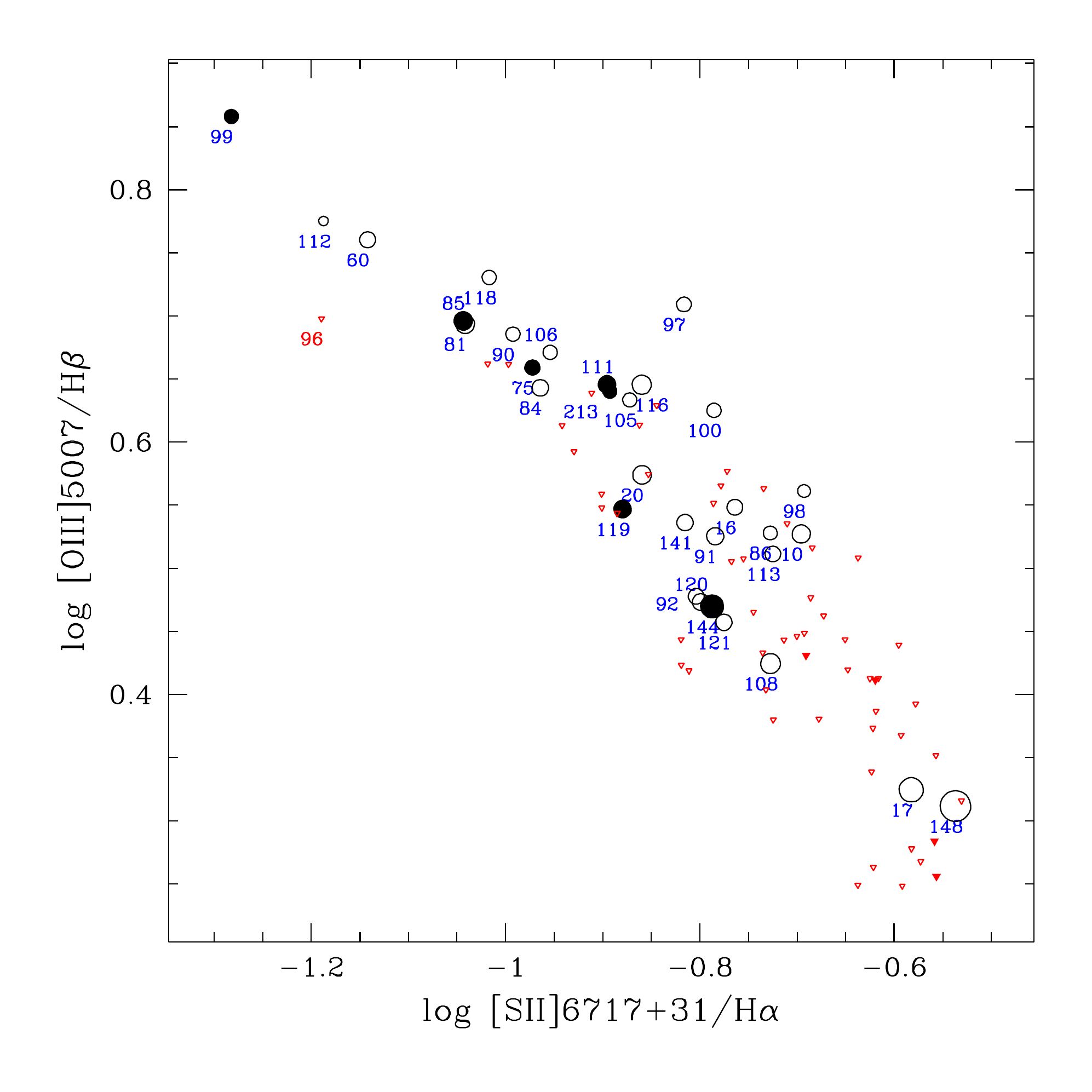}
\includegraphics[width=0.495\linewidth]{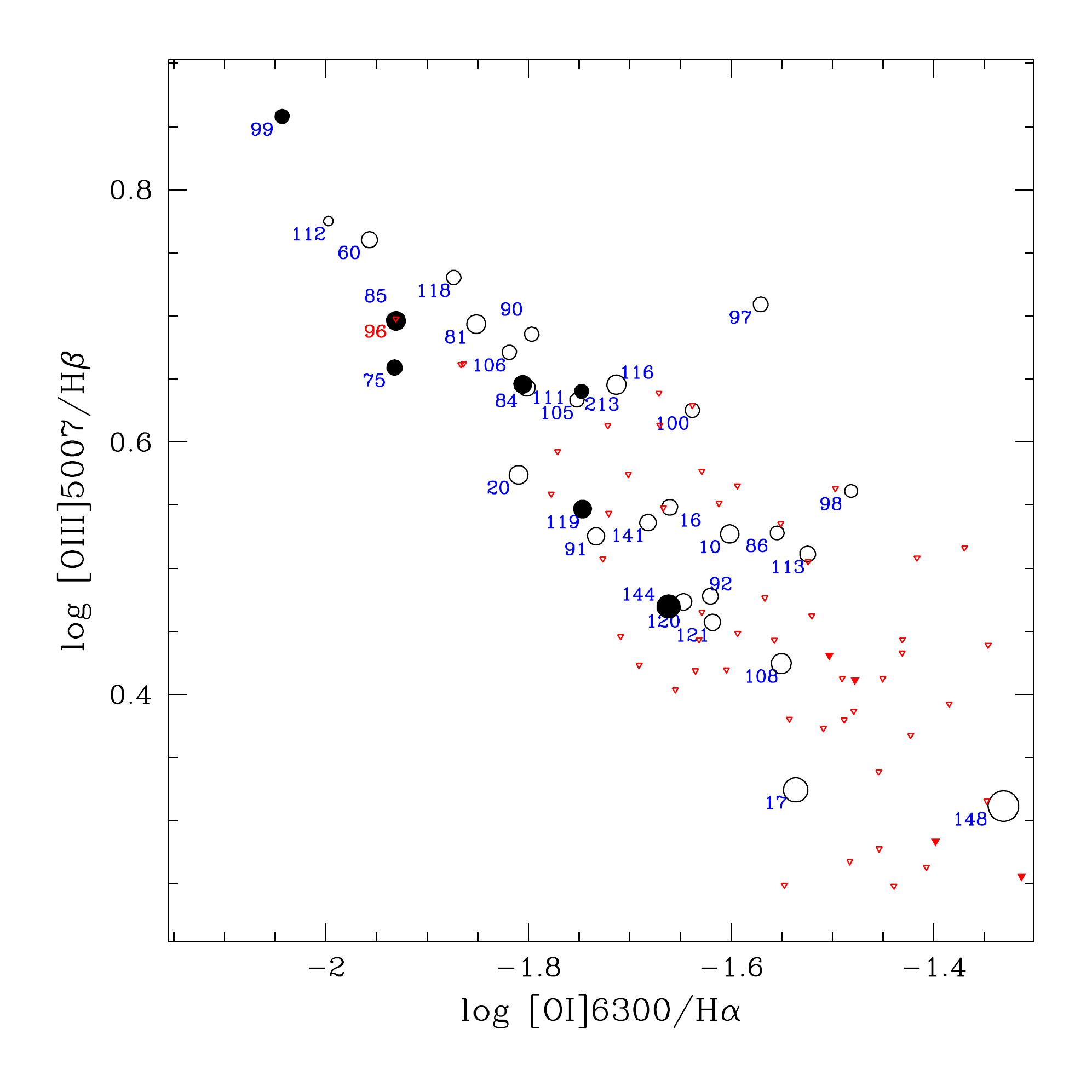}
\includegraphics[width=0.495\linewidth]{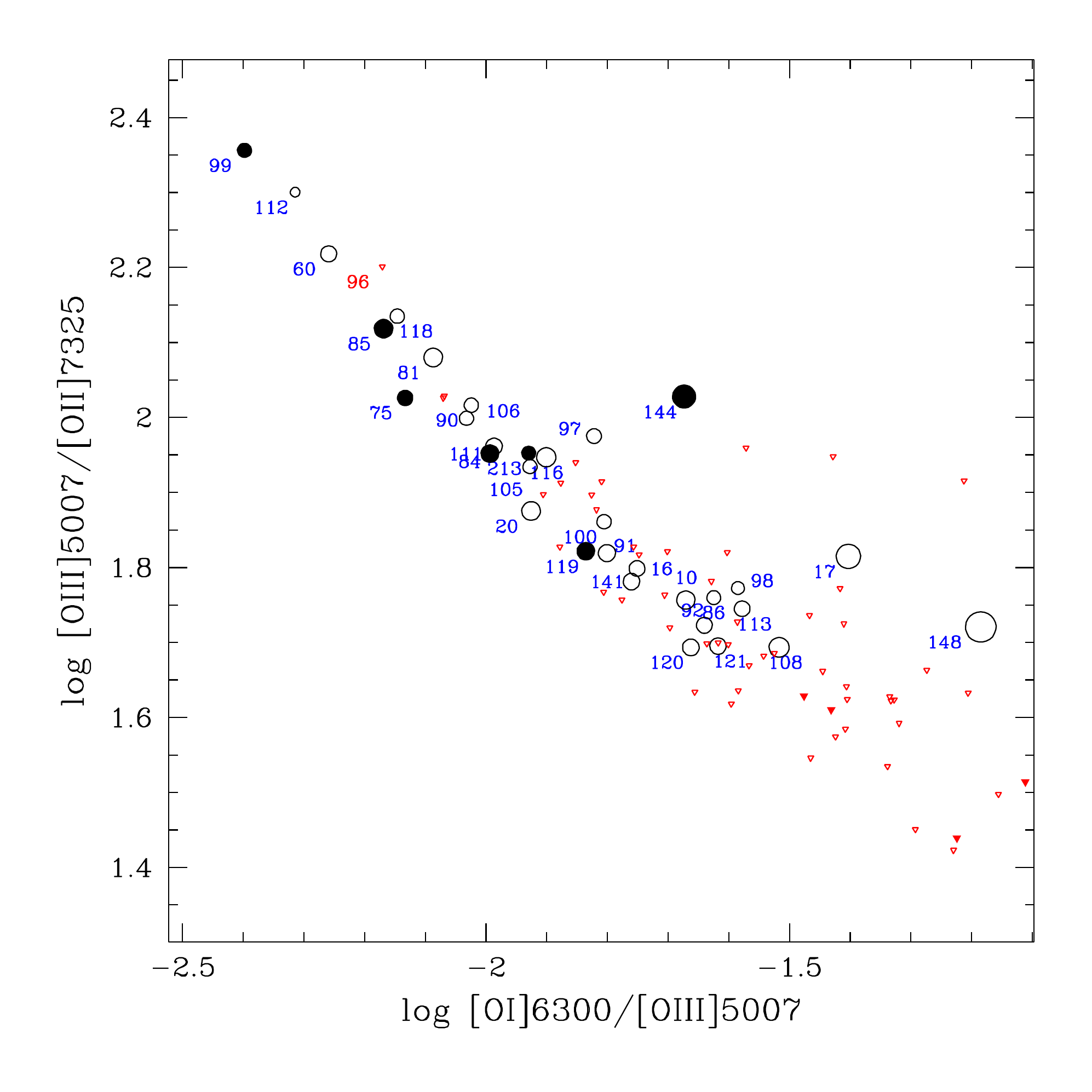}
\par\end{centering}
\caption{
Cartwheel \hii\ regions in BPT diagrams following the same symbol convention as in Figure~\ref{fig:heii_ew}; (upper left) \oiiibyhb\ vs. \niibyha\ ratio; (upper right) \oiiibyhb\ vs. \siibyha\ ratio; (bottom left) \oiiibyhb\ vs. \oibyha\ ratio; (bottom right) \oiiibyhb\ vs. \oibyoiii\ ratio. See text for details.
}
\label{fig:heii_bpt}
\end{figure*}

\begin{figure}
\begin{centering}
\includegraphics[width=1.0\linewidth]{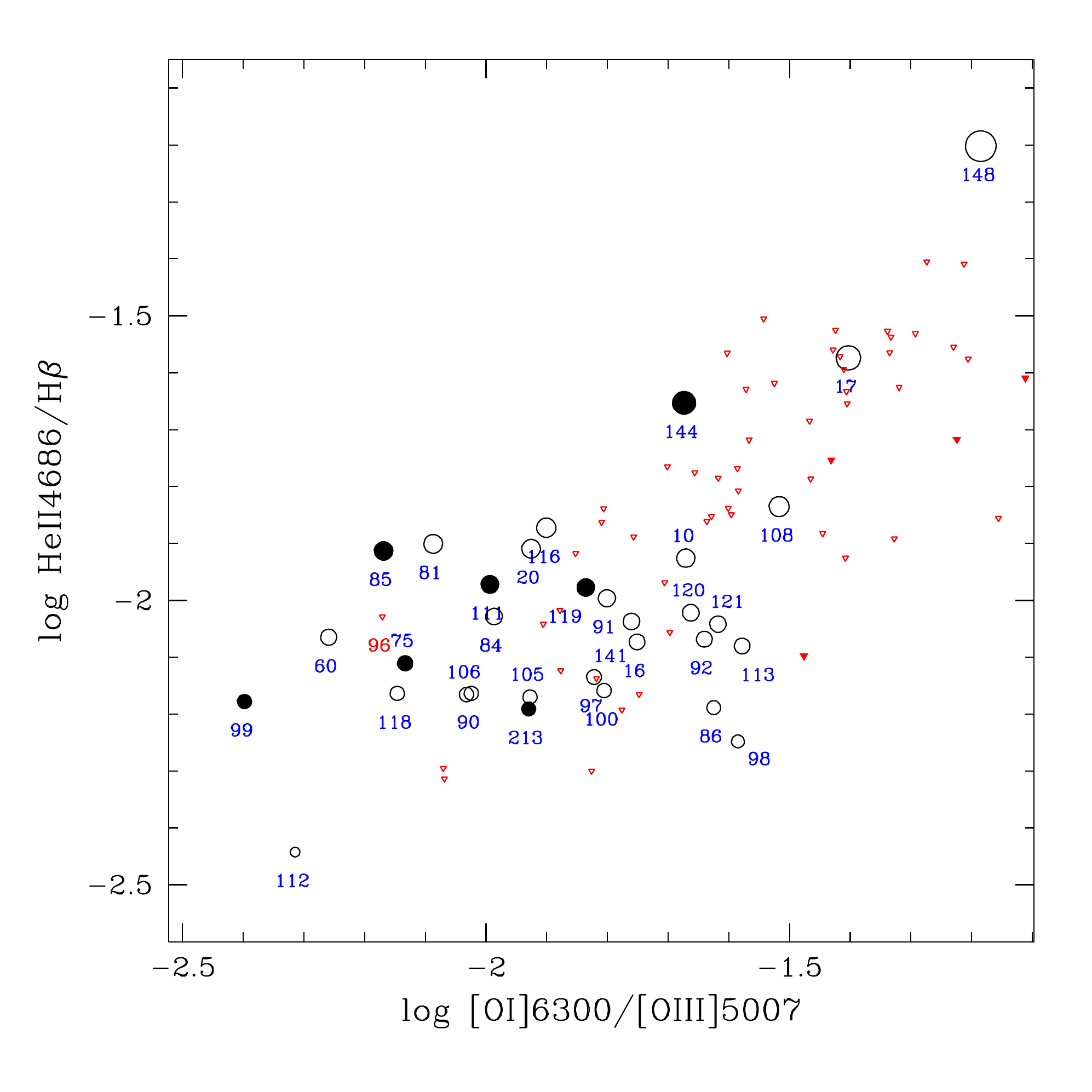}
\par\end{centering}
\caption{The \heiibyhb\ ratio plotted against the shock-sensitive \oibyoiii\ line ratio. The meaning of the symbols are the same as in Figure~\ref{fig:heii_ew}.
}
\label{fig:heii_shock}
\end{figure}

We start our analysis of the diagnostic line ratios by plotting in Figure~\ref{fig:heii_ew}
the \heiibyhb\ ratio versus \ewhb, which is a standard indicator of age of 
stellar populations during their early nebular phase, as illustrated in Figure~\ref{fig:EWHb_evol1}. 
In this and all the upcoming figures, we distinguish \hii\ regions with and
without the detection of the \heiiwr\ nebular line by circles and inverted triangles, respectively.
Filled symbols show the \hii\ regions that have an associated ULX. 
Error bars are shown only when the errors are significantly
larger than the symbol size. \hii\ regions with detected \heiiwr\ line are
annotated with their number designation from Table~\ref{tab:heii_regions}. 
Care is taken to avoid superposition of the annotations, but it was
not always possible due to the crowding of points in some of the plots.

Figure~\ref{fig:heii_ew} shows that the \heiiwr\ line is detected in seven of the 12 \hii\ regions with \ewhb$>$100~\AA. Among the high \ewhb\ regions without \heiiwr\ detection, region\#96 is identified in the plot. This region lies slightly outside the ring in the bright southern arc (see Figure~\ref{fig:zoom_image}). The \hii\ regions with lower emission \ewhb\  have, in general, fainter nebular lines making the detection of the faint \heiiwr\ line dependent of the SNR of each spectra (see Figure~\ref{fig:heii_snr}). Three regions (\#144, \#17 and \#148) standout in the diagram, as they are among the regions with the lowest \ewhb, but having the highest values of \heiibyhb\ ratio. In fact, these three regions exemplify a tendency for the upper boundary of the \heiibyhb\ ratio to increase with decreasing \ewhb.

We show the \oiiibyhb\ ratio against \ewhb\ on the left panel of the 
Figure~\ref{fig:heii_oiii}. The \oiiibyhb\  ratio is a well-known indicator of the ionization 
state of an \hii\ region, with high ionization regions having a higher value 
of the ratio. We clarify that the error bars on the ratio are negligibly small, 
including for those regions without the \heiiwr\ detection (inverted triangles).
Note that the higher \ewhb\ regions have higher \oiiibyhb\ ratio, independent of whether the \heiiwr\ line is detected or not, with region \#99 (the brightest \hii\ region) and \#148 
(\hii\ region with the highest \heiibyhb\ ratio) lying at the extreme ends of the
relation shown by the rest of the regions. The region \#111 stands out from the relation for having a too high ionization for its observed low \ewhb. We recall that this source is the \hii\ region nearest to the HLX source. However, its association with the HLX source is doubtful as discussed in Sec.~\ref{sec:ulx}. Dilution of EW(\hb) from a non-ionizing cluster inside the aperture used for extracting the spectrum is the most likely reason for this region to displace from the observed correlation. This is supported by a visual examination of the HST images of this region, which reveals two sources, with the source brighter in the F814W image the likely non-ionizing cluster.

In the right panel, we show the \heiibyhb\ ratio against the \oiiibyhb\ ratio. 
As expected, \heiiwr-line detection is more frequent in high ionization \hii\ regions as compared to relatively low ionization regions --- it is detected in as much as 75~per cent (15 out of 21) of the high ionization regions (log(\oiiibyhb)$>$0.60). Surprisingly, low ionization \hii\ regions (log(\oiiibyhb)$<$0.40) have a non-zero detection frequency (10~per cent; two out of 20). 
The increasing tendency for the upper boundary of \heiibyhb\ value with decreasing \ewhb\ is also manifested in the \heiibyhb\ vs \oiiibyhb\ plot, with \#99 and \#148 marking the endpoints of this tendency.

\subsection{Location of \heii-emitting \hii\ regions in diagnostic diagrams}

In order to understand the sources of ionization of He$^+$ in \hii\ regions of 
Cartwheel, we show in Figure~\ref{fig:heii_bpt} all our regions in the most commonly used 
BPT diagrams \citep{BPT1981}. In the first three plots (top two and the bottom left), we
use lines of low ionization potential in the x-axis, whereas the y-axis
contains \oiii\ line, which as discussed before originates in the high ionization zone. 
The \hii\ regions lie along a sequence wherein the ratios of \niibyha, \siibyha\
and \oibyha, systematically increase as the \oiiibyhb\ ratio decreases. The fraction of \heii-emitting regions decreases along the sequence from top-left to bottom-right.
In the bottom right panel, we show line ratios that maximize the values for high ionization regions on the y-axis and low ionization regions on the x-axis. In this plot, the relation is much tighter than in other plots with the \heii-emitting regions having the lowest value of \oibyoiii\ ratio for any fixed \oiiibyoii\ value.

In order to investigate the possible presence of shock ionization/excitation in \hii\ regions having \heiiwr\ lines, in Figure~\ref{fig:heii_shock} we plot the \heiibyhb\ against \oibyoiii\ ratio, which is sensitive to the presence of shocks \citep[see e.g. Figure~15 in][]{Plat2019}. The four regions with the highest values of \heiibyhb\ (\#148, 17, 144 and 108) are indeed among the \hii\ regions with the highest values of the shock-sensitive ratios.

The regions ionized by the ULX sources are expected to have \oiiibyhb$>5$ and \oibyha$>$0.1, occupying the transition region between the Active Galactic Nuclei (AGNs) and Low-Ionization narrow-emission line regions (LINERs) in the BPT diagrams \citep{Gurpide2022}. None of the ULX sources in the Cartwheel occupy these zones, suggesting that the ULX sources have very limited role, if any, in the ionization of the nebula with which the ULX source is positionally coincident.

\section{Discussion}

We now investigate the source of He$^+$ ionization in the Cartwheel using
the trajectory of theoretical models of ionization in various diagnostic diagrams using the theoretical line ratios calculated by \citet{Plat2019}.
In particular, we use the diagnostic diagrams involving \oiiibyhb\ ratio plotted against
\niibyha\ and the \ewhb, \oiiibyoii\ vs. \oibyoiii, \heiibyhb\ vs. \ewhb, \heiibyhb\ vs \oibyoiii\ and \oiiibyoii\ vs \arivbyariii. These diagrams are chosen because they are sensitive to the different mechanisms of ionization explored in this work.

\subsection{Calculation of theoretical nebular line ratios}

We consider two sources of ionization: photoionization, which is the most 
dominant source of ionization in \hii\ regions, and ionization by radiative shocks. 
Photoionization by stellar clusters with and without binaries is considered.
Since the objects under study are \hii\ regions, the ionizing source is better modelled as a single age cluster (SSP), rather than a population of stars formed over a long period of time. 
We hence consider only SSP models. However, the dilution of EW(\hb) caused by the presence of any non-ionizing source (e.g. an underlying old stellar population) inside the aperture used for extraction is taken into account.

The emission line fluxes from an \hii\ region are computed with CLOUDY v17.02 \citep{Ferland2017} following the approach of \citet{Gutkin2016}. We use C\&B for single star models and BPASS v2.2.1 for binary population models. 
The stellar and interstellar medium (ISM) metallicity is set to $Z=0.003$, which corresponds to the gas phase oxygen abundance of $12+\log{\rm (O/H)}_{\rm gas}\approx 8$ for the dust-to-metal mass ratio, $\xi_{d}$=0.3 (the solar value is 0.36) following \citet{Gutkin2016}. The ISM is considered to be of uniform hydrogen density of $n_{\rm H}=10^2\,$cm$^{-3}$, which along with the rate of ionizing photons and filling factor sets the ionization parameter $U$. Models are parametrized in terms of the zero-age volume-averaged ionization parameter $\langle U\rangle$, which is varied by varying the filling factor, see \citet{Gutkin2016} for details. 
 We note that the values of $12+\log{\rm (O/H)}_{\rm gas}$, $\log{\rm (N/O)}_{\rm gas}$ and $n_{\rm e}$ used here are based on  the mean values derived using the same MUSE dataset for the Cartwheel ring \hii\ regions in a companion paper \citep{Zaragoza2022}. 

The \hii\ regions are assumed to be ionization-bounded, but contain dust grains that compete with gas in the absorption of ionizing photons. We also discuss the effect on the line ratios if the \hii\ regions are density bounded, i.e. when the size of the \hii\ region is limited by the gas density, rather than the ionizing photon rate. We account for the possible presence of holes or cavities in the \hii\ regions by means of an escape fraction of ionization photons. In addition to these two escape geometries, we follow the approach of \citet{Ramambason2020} and compute two-zone models, combining a low and high ionization parameter component. These two zones are either both ionization bounded, or one of them is density bounded.

The emission from fast radiative shocks is added using the models of \citet{Alarie2019}\footnote{The ISM metallicity in the shock models corresponds to the SMC metallicity, which is slightly lower than that used for our \hii\ regions. The nitrogen to oxygen abundance ratio is also lower in these models as compared to that used in our \hii\ region models.} for the full set of shock velocities (100 to 1000~km\,s$^{-1}$) available in these models. A pre-shock density of $10^2\,$cm$^{-3}$ and transverse magnetic field strength $B=1\mu G$, which are typical values for the diffuse ISM in galaxies, are used. The calculated line ratios depend weakly on these fixed values, as compared to the variation in shock velocities. The effects of shocks are added at representative ages, which allows us to illustrate the effect of shocks on line ratios as the cluster evolves. 

\begin{figure*}
\begin{centering}
\includegraphics[width=1\linewidth]{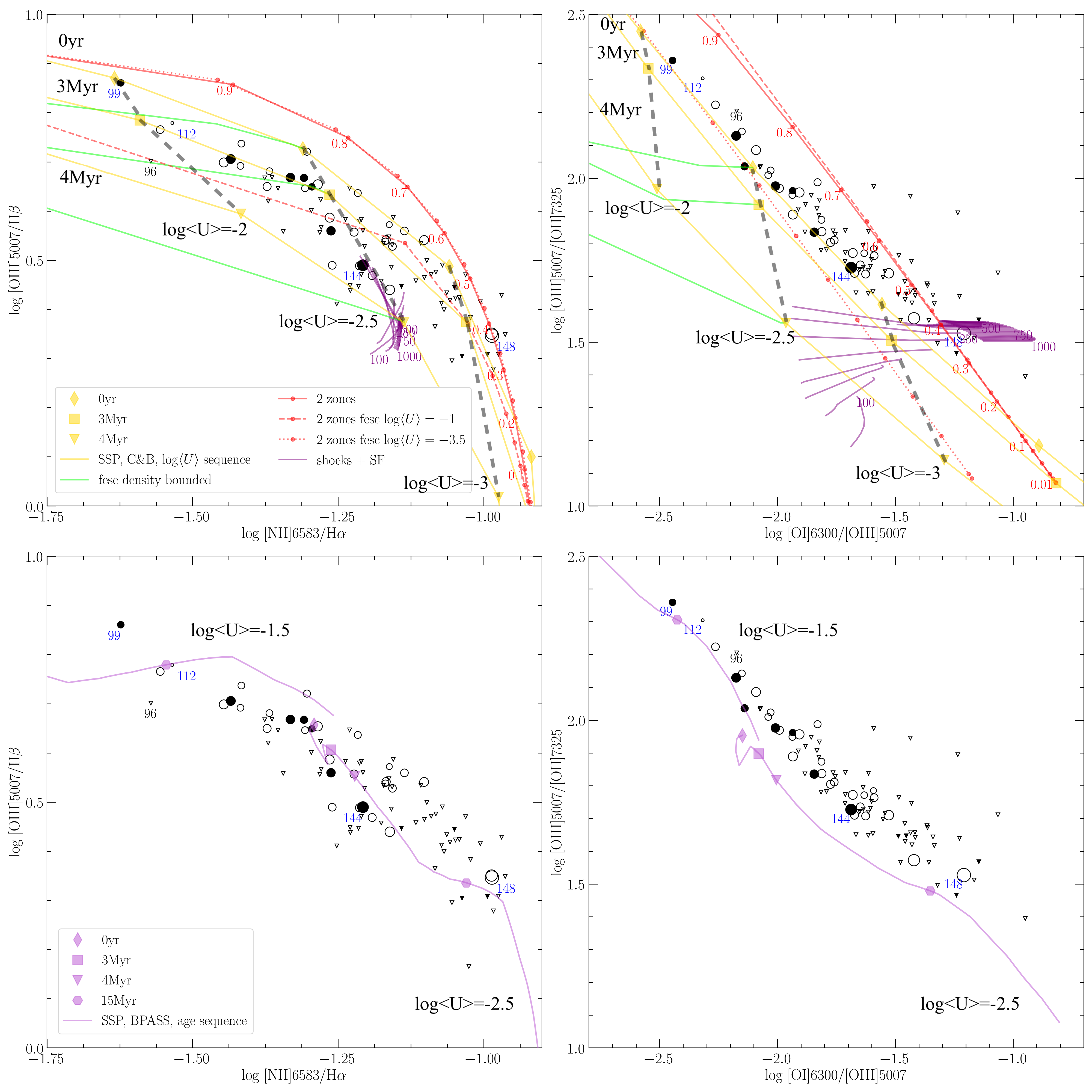}
\par\end{centering}
\caption{Theoretical models superposed on the observed points of the Cartwheel in the line ratio diagnostic diagrams illustrated in Figure~\ref{fig:heii_bpt}.
Top panels: C\&B model sequences, formed by varying the initial \logU\ from $-1$ to $-4$, at 0, 3 and 4~Myr (solid yellow lines). The evolutionary tracks at \logU=$-2$, $-2.5$ and $-3$ are shown by black dashed lines. Effect of regions being density-bounded (solid green lines at \logU=$-2.5$), two-zone escape (red lines joining \logU=$-1$ and $-3.5$ at 3~Myr), and shock models (purple tracks at 4~Myr and \logU=$-2.5$) are illustrated at selected locations in the diagrams. The numbers along the two-zone model sequences correspond to the fractional contribution to the \hb\ flux from the high \logU\ zone, and the numbers in the shock models correspond to shock velocities that range from 100 to 1000~km\,s$^{-1}$.
Bottom panels: BPASS binary SSP age sequences for two values of initial \logU. 
See the inset for identification of various lines and symbols, and Sec.~4.2 for more details of the models used in these plots.
}
\label{fig:BPT1}
\end{figure*}

\subsection{The line ratio sequence from photoionization models}

\begin{figure*}
\begin{centering}
\includegraphics[width=1\linewidth]{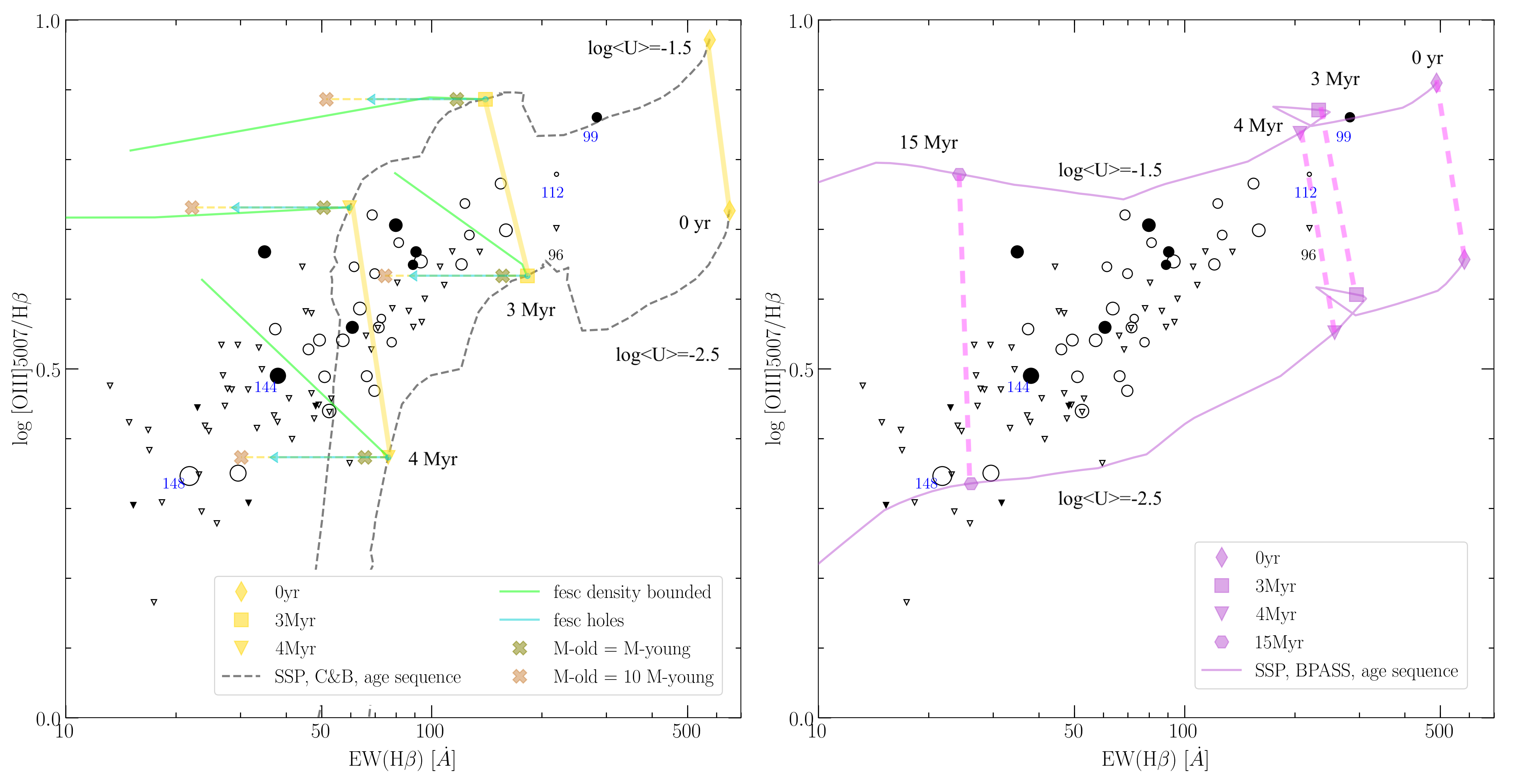}
\par\end{centering}
\caption{
Comparison of the observed values for the Cartwheel \hii\ regions (circles and small inverted triangles; see caption of  Figure~\ref{fig:heii_ew} for their meaning) with theoretical models. Trajectory of evolutionary tracks for evolving stellar populations at two values of initial \logU=$-1.5$ and $-2.5$ are shown (dashed black lines) with the points on the two tracks at selected ages joined by yellow lines. The left panel shows the C\&B SSP tracks without binaries, where we also show the effect of escape of ionizing photons for two escape geometries (holes and density bounded) at ages 3 and 4~Myr, and the effect of adding a non-ionizing cluster of 10~Myr that is 1 and 10 times more massive than the ionizing cluster. In the right panel, we show the BPASS SSP models including binary stars. 
}
\label{fig:oiii_new}
\end{figure*}

In Figures~\ref{fig:BPT1} and \ref{fig:oiii_new}, 
we overplot tracks for photoionization models for different initial $\log \langle U\rangle$ values using C\&B SSP models without binary stars, and BPASS models with binary stars.
The ratios for leaky \hii\ regions, as well as for \hii\ regions
experiencing radiative shocks are also explored. Additionally, the effect of an 
underlying old population is illustrated.
We also explored the effect of binaries using BPASS models. We refer the reader to \citet{Plat2019} for a detailed illustration of the sensitivities of the explored parameters on the commonly observed optical and ultraviolet line ratios.
In different plots, we plot sequences of age, or $\log \langle U\rangle$ or shock velocities,
depending on the sensitivity of the plotted quantities on the models. 
These are explained in annotations and legends of the corresponding figures.

\begin{figure*}
\begin{centering}
\includegraphics[width=1\linewidth]{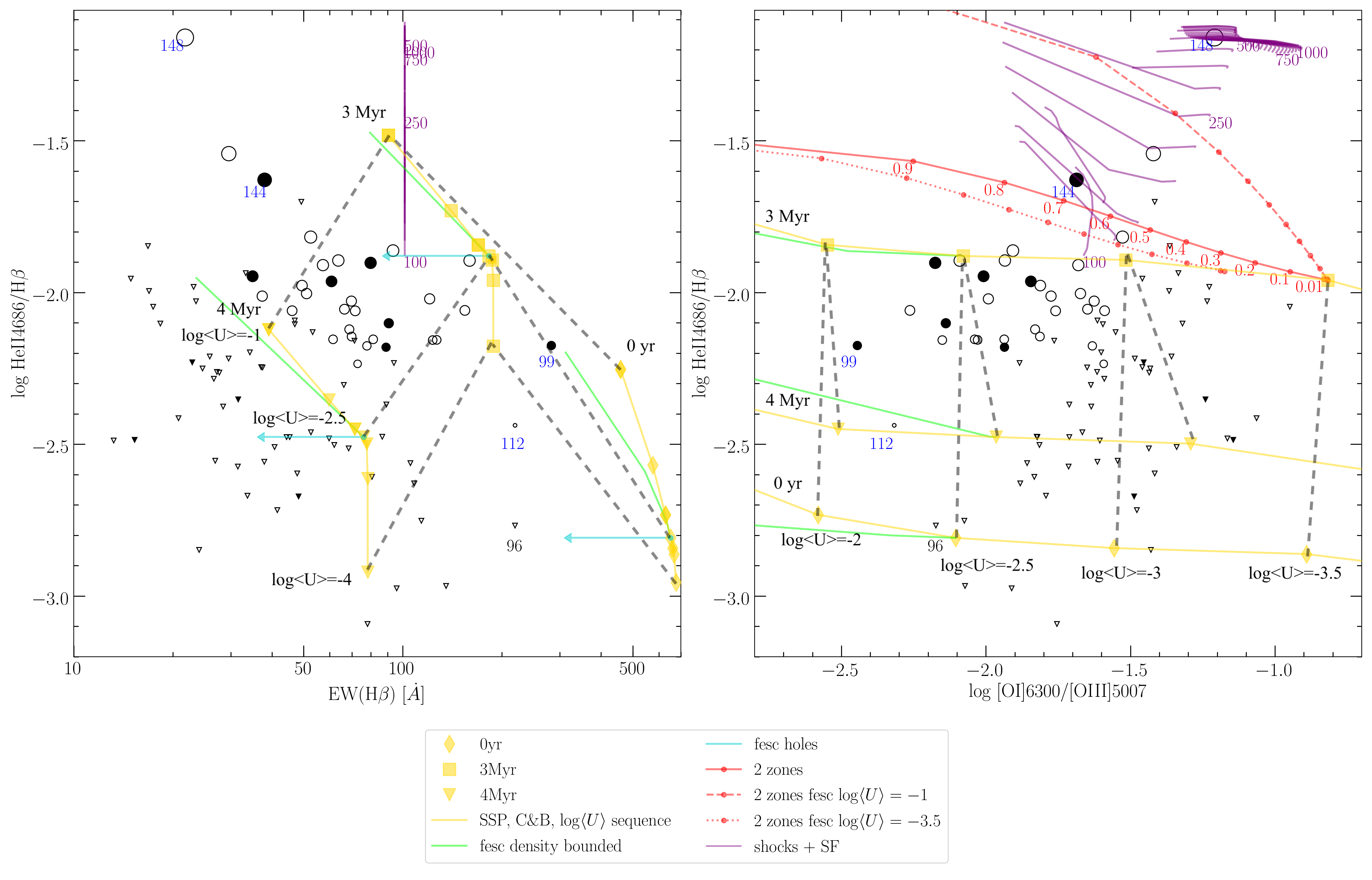}
\par\end{centering}
\caption{
(left) \heiibyhb\ vs. \ewhb; (right) \heiibyhb\ vs. \oibyoiii\ ratio. See the guide to the lines and symbols in the box below the graphs and the text in the beginning of Sec.~4.2 for the explanation of parameters for the shock and 2 zone ionization models, and see caption of Figure~\ref{fig:heii_ew} for the meaning of symbols for the observed points (circles and small inverted triangles).
}
\label{fig:heii_mod}
\end{figure*}

The cluster evolution is followed up to 50~Myr with three epochs (0, 3, and 4~Myr) marked with differently shaped symbols (in some of the plots, the plotted range covers only the trajectory around 3--4~Myr). The initial $\log \langle U\rangle$ between $-1$ and $-4$ are explored. In shock models, shocks are assumed to provide 25~per cent of the observed \hb\ flux. The sequences are formed for shock velocities between 100 to 1000~km\,s$^{-1}$. High velocity shocks are expected in \hii\ regions following the explosion of SN, which start occurring at an SSP age of $\sim$4~Myr. Hence, the shock component is added to the 4~Myr \hii\ region emission with $\log \langle U\rangle =-2.5$. These are shown in Figures~\ref{fig:BPT1}, \ref{fig:heii_mod} and \ref{fig:Ar} by purple lines. In models involving escape of ionizing photons, the sequence is formed by varying the fraction of Lyman continuum (LyC) photons escaping the \hii\ regions. In two-zone models, the sequence is formed by varying the fractional contribution to the \hb\ flux from the high \logU\ zone. These are shown in Figures~\ref{fig:BPT1}, \ref{fig:heii_mod} (right) and \ref{fig:Ar} with red lines marked with dots for increase of the escape fraction between 0.1 and 0.9.

We now discuss the results on the ionization mechanisms suggested by the models based on the comparison of the locus of model parameters with observations, in selected line-ratio diagrams. We first discuss the results based on C\&B models without binary stars, and at the end, comment on the effect of having binary stars in the SSPs.

{\it Effect of \logU:}
The density of the ISM surrounding the cluster, the rate of ionizing photons, and the volume filling factor of the ionized gas fix the initial \logU\ of the models. In the top-left panel of Figure~\ref{fig:BPT1}, it can be seen that the observed sequence of \oiiibyhb\ values can be understood as due to a dispersion in the initial \logU\  values, with the ionizing clusters in almost all of the \hii\ regions having ages between 3 and 4~Myr in C\&B models. The observed range of line ratios is covered by the models with \logU=$-1.5$ to $-3.0$, with the small variations in the age being responsible for the observed spread in the direction perpendicular to the sequence. The observed sequence in the \oiiibyoii\ vs \oibyoiii\ plane  (top-right) is also consistent as a sequence of initial \logU. However, the inferred age from this  diagram is $\sim$0~Myr, rather than 3--4~Myr inferred from the \oiiibyhb\ vs \niibyha\ diagram. The latter diagram is sensitive to the selective escape of photons from either the low or the high ionization zones as will be discussed in the two-zone models below.

{\it Effect of dust in \hii\ regions:}
All our models include dust inside the \hii\ regions. The quantity of dust is parametrized by the dust-to-metal mass ratio $\xi_{\rm{d}}$. The dust competes with gas in the absorption of ionization photons following the scheme proposed by \citet{Bottorff1998}.  Hence the number of ionizing photons absorbed by hydrogen, and all properties that depend on the ionizing photon flux, predicted by our models is lower than those predicted for dust-free ionization-bounded models. The optical depth of ionizing photons arising from dust is proportional to the total hydrogen column density, which is proportional to the ionization parameter. So as the ionization parameter increases, so does the absorption of ionizing photons by dust rather than hydrogen. This effect can be noticed in Figures~\ref{fig:oiii_new} and \ref{fig:heii_mod}, where the \ewhb\ decreases with \logU\ at a fixed age \citep[see also][]{Erb2010,Plat2019}.

{\it Effect of upper cut-off mass of the IMF:}
The SSPs we used for the calculation of nebular quantities correspond to $M_{\rm u}$=300~\msol. The non-detection of BB (see Sec.~2.6.1) in our spectra suggests the absence of stars more massive than 100~\msol\ in Cartwheel \hii\ regions. However, the results for $M_{\rm u}$=100~\msol\ are identical to that for $M_{\rm u}$=300~\msol\ after 3.2~Myr, as can be inferred from Figures~\ref{fig:Nwr_CB_evol} and \ref{fig:Nwr_vs_QHeII}. The expected values of \heiibyhb\ before and during the WR phase with $M_{\rm u}$=300~\msol\ is only marginally higher than that for $M_{\rm u}$=100~\msol. 
Thus, the results presented here are not sensitive to the choice of upper cut-off mass as long as the SSP contains hot massive-stars that go through the WR phase ($M_{\rm u}>$25~\msol).

{\it Effect of cluster evolution:}
The rate of ionizing photons emanating from a cluster starts decreasing when the 
most massive stars, also the hottest and the most luminous, end their main sequence 
life time. For clusters with the highest mass $\sim$100~\msol, this starts happening
at $\sim$3~Myr at Z=0.003, the metallicity corresponding to the Cartwheel. 
This decreases  \logU\  by $\sim$0.6~dex over the first 10~Myr, which leads to a gradual
decrease of the high ionization line intensity ratios such as \oiiibyhb.
The \ewhb\ decreases slowly in the first 3~Myr reaching values $\sim$200~\AA\ at 3~Myr, 
beyond which it drastically drops by more than an order of magnitude in $\sim$10~Myr.
The decrease of \logU\  and the \ewhb\ with age leads to a decrease of \oiiibyhb\ ratio as the \ewhb\ decreases.

The evolutionary track in C\&B models for a given initial \logU\ in Figure~\ref{fig:oiii_new} follows the observed relation for clusters younger than $\sim$4~Myr beyond which the model-predicted \oiiibyhb\ ratio is much smaller than that expected from the observed relation. Consequently, \hii\ regions with \ewhb$<$50~\AA\ are not expected to have detectable levels of the \oiii\ line emission. The continuity of the observed sequence in this diagram for the whole range of \ewhb\ suggests that the Cartwheel \hii\ regions are indeed ionized by clusters younger than 4~Myr, and some physical effect is responsible for lowering the observed values of \ewhb\ compared to that predicted in the SSP models we used. The ionization sequence in Figure~\ref{fig:BPT1} also implies that all our \hii\ regions are younger than 4~Myr. Escape of ionizing photons, presence of density-bounded \hii\ regions, presence of an older underlying population, are some of the physical processes that we have explored in this work to explain the decrease of \ewhb\ without the corresponding decrease in \oiiibyhb\ ratio.

{\it Effect of an older population:} 
\hii\ regions often contain a population other than that is ionizing the surrounding gas inside the aperture used for spectral extraction\citep[e.g.][]{Charlot1993, Mayya1996}. The non-ionizing population could be a cluster older than the ionizing cluster, or it could be the underlying disk population. Given the recent star formation history of the Cartwheel, it is likely that the apertures used for spectral extraction (740~pc diameter) contain non-ionizing clusters of $\sim$10~Myr or slightly older. Results from the recent numerical simulation of the wave propagation in the Cartwheel by \citet{Renaud2018} support the existence of a spread of this order in the ages of clusters in the outer ring. Multiple clusters within the extracted apertures can be directly seen in Figure~\ref{fig:zoom_image} at the spatial resolution of the HST images.
The presence of such a cluster would decrease the \ewhb\ without affecting the line ratios, and hence would move the points horizontally in Figure~\ref{fig:oiii_new}. 
The \ewhb\ would be affected by a larger amount for larger mass of the non-ionizing cluster. In the left panel, we show the effect of a non-ionizing cluster of 10~Myr age for two values of relative masses: (1) equal masses, and (2) the non-ionizing cluster 10 times more massive than the ionizing cluster. This effect is shown by crosses placed at 3 and 4~Myr of age for the ionizing cluster. A part of the observed horizontal spread could be due to the presence of different amounts of mass in old stellar clusters inside the apertures used for extraction.

{\it Effect of escape of ionizing photons:} We have assumed that all the ionizing photons that are emitted by the clusters are either used in the ionization, or absorbed by internal dust. However, there are 2 geometries by which ionizing photons can escape the nebula without
getting absorbed by gas and dust: (1) escape through holes, and (2) escape through density-bounded zones. The former case results in the decrease of the intensity of all lines, hence a decrease in the \ewhb, without changing the line ratios. Thus, the escape of ionizing photons through holes would move the points horizontally in Figure~\ref{fig:oiii_new}, producing an effect indistinguishable from the presence of a non-ionizing population discussed above. The second case is discussed below.

{\it Density bounded models:}
\hii\ regions have an ionization structure with the lines of high ionization (e.g. \oiii) 
originating in zones closer to the ionizing cluster as compared to the lines of low 
ionization (e.g. \nii$\lambda$6583). In density-bounded \hii\ regions, the ionizing
photons are lost due to insufficient amount of gas to trap all the ionizing photons.
Such regions would have a reduced intensity of low-ionization lines and \ewhb\ as compared to
the values for ionization-bounded regions. This would lead to an increase of \oiiibyhb\ as
\ewhb\ decreases, which is just the opposite of what is observed in Figure~\ref{fig:oiii_new}.
Hence, density-bounded models cannot explain the continuation of the sequence to low \ewhb.
Furthermore, the observed sequence in \oiiibyoii\ vs \oibyoiii\ ratio is not consistent with density-bounded models.

{\it Two-zone escape models:}
In order to reproduce the observed large range of \oibyoiii\ ratios in LyC galaxies,
\citet{Ramambason2020} proposed a model wherein the nebulae loose ionizing photons 
selectively from the high-$U$ or low-$U$ zones. Such a configuration is naturally expected if the ISM around the clusters is non-uniform, and contains dense clumps and filaments. Results for two-zone models are calculated by combining the spectrum of a high initial \logU\ nebula with a second spectrum of a low initial \logU\ nebula, with the relative luminosity of the \hb\ lines in the two spectra used as a free parameter. This free parameter, $\omega$, varies between 0 and 1, which is defined as the fractional contribution of the high \logU\ spectrum to the total \hb\ luminosity. Thus, $\omega$=1 corresponds to a nebula containing only high ionization zone and $\omega$=0 corresponds to a nebula containing only low ionization zone. The ionization-bounded or density-bounded status of the two zones are independent of each other. In Figure~\ref{fig:BPT1} (right), we show the locus of these models by red lines for three cases, all corresponding to 3~Myr age clusters with initial \logU=$-1$ and \logU=$-3.5$ for the high and low ionization zones, respectively. The 3 cases are:  (1) both the zones are ionization-bounded (solid line),  (2) combination of an ionization-bounded low \logU\ zone, with a density bounded high \logU\ zone with an escape fraction, $f_{\rm esc}\sim30$~percent (dashed line), and  (3)  combination of an ionization-bounded high \logU\ zone, with a density bounded low \logU\ zone with $f_{\rm esc}\sim10$~percent (dotted line). Values for the free parameters are chosen that best illustrate the observed trends in Figure~\ref{fig:BPT1} (right).
C\&B ionization-bounded models (yellow lines) even for the youngest age lie slightly to the left of the observed points in this figure. The tracks for ages of 3 and 4~Myr are further away from the observed points. 
On the other hand, the two zone escape models are able to explain the dataset for cluster ages of 3~Myr --- most of the observed points lie between the dashed red lines (3~Myr track with selective escape of ionization photons from high \logU\ zone) and the yellow line for the 3~Myr track (ionization bounded regions without any escape).
Thus, with two-zone models, it is possible to obtain consistent ages of $\sim$3--4~Myr for the majority of the \hii\ regions in the Cartwheel, from both the diagnostic diagrams in Figure~\ref{fig:BPT1}. 

{\it Shocks:}
Some of the \hii\ regions, though principally photoionized by the stars, can also
experience shocks, especially during the post-main sequence evolution and the death of massive
stars. In order to illustrate the effect of shocks in the line ratio diagrams, shock models of \citet{Alarie2019} are added to the evolutionary track of SSP models at one particular epoch (4~Myr) with $\log \langle U\rangle=-2.5$. The shocks are parametrized by 3 parameters: the percentage of energy in shocks compared to photoionization energy to ionize hydrogen, the velocity of the shock and the shock age. 
We used their grid of truncated shock models. We here plot models where 25~percent of the total \hb\ flux is provided through shocks for various values of shock velocities. At each velocity, we plot the ratios as the shock propagates into the ISM.
Increasing the shock velocity increases the line ratios involving high ionization, whereas the low ionization lines become stronger as the shock propagates.
The effect of shocks on the line ratios can be best seen in the \oiiibyoii\ vs. \oibyoiii\ diagram of Figure~\ref{fig:BPT1} (top right panel). High velocity shocks tend to move the points to the right, mostly occupied by \hii\ regions not showing \heiiwr\ line. Among the regions with \heiiwr\ emission, shocks could be important in low-ionization \hii\ regions such as \#148. On the other hand, shocks have very little effect in the \oiiibyhb\ vs \niibyha\ plot (top left panel), and no effect on the \ewhb.

{\it Cluster evolution with binaries:}
We now discuss the effect of binary population on the results obtained from C\&B models.
For this purpose, we use the BPASS binary models plotted in the bottom panels of Figures~\ref{fig:BPT1}, and right panel of Figure~\ref{fig:oiii_new}. The result that the observed sequence of points is a sequence in \logU\ holds even after including binary stars in the SSPs. 
Star clusters containing binary stars produce the ionizing photons over an extended 
period of time ($>$15~Myr) as compared to the evolution without binaries. This 
causes both the \oiiibyhb\ and the \ewhb\ to decrease more slowly with evolution as compared to the evolution of clusters without binary stars. The evolutionary locus and \logU\ sequences in binary models are almost parallel in Figure~\ref{fig:oiii_new}, both following the ionization sequences seen in these diagrams. Thus, under the binary models, Cartwheel \hii\ regions could have an age range between 0 and 15~Myr and \logU\ range between $-1.5$ to $-3.5$. However, at ages as late as 15~Myr, the EW(\heiiwr) decreases by more than an order of magnitude (see Figure~\ref{fig:EWHb_evol1}). Hence, it is highly unlikely that the \heiiwr\ detections correspond to the faint late phase, rather than the luminous early phase. 

Thus, the inclusion of binary models does not qualitatively change the conclusions arrived from using C\&B models, which do not take into account the possible presence of binary stars.

\subsection{On the ionization state of Cartwheel \hii\ regions}\label{sec:ionization_state}

The observed range of line ratios in the \hii\ regions of the Cartwheel 
corresponds to photoionization by young clusters (age$\sim$3--4~Myr in C\&B models or 3--15~Myr in BPASS binary models) with initial \hii-averaged ionization parameter \logU\  lying between $-1.5$ to $-3.5$. The \oiiibyhb\ ratio, which is a direct measure of the degree of ionization of an \hii\ region, is  correlated with \ewhb\ 
over the entire observed range of both the quantities. This correlation suggests a systematic decrease of the \oiiibyhb\ ratio, or equivalently \logU, with age. For the ionization-bounded models that we have used, \logU\ decreases only by 0.6~dex or alternatively \oiiibyhb\ by 0.2~dex in 4~Myr in C\&B SSP models and 0.1~dex in 15~Myr in BPASS binary SSP models. The amount of change of this ratio in SSP models depends on the metallicity, with the values in our models being consistent with the values obtained by \citet{Stasinska1996} for the Cartwheel metallicity. Thus, an age-dependent process of softening of \logU\  is required in order to interpret the observed correlation between \oiiibyhb\ and \ewhb\ primarily as an age sequence. 

\citet{Kim2019} found that the fraction of the escape of the ionizing photons systematically increases with age of the cluster due to the increased amount of feedback with age. 
The escape of ionizing photons has been long suspected to be one of the reasons for the low \ewhb\ of \hii\ regions \citep[e.g.][]{Mayya1996}. \citet{Castellanos2002} determined escape fraction between 0.1--0.7 for three \hii\ regions they analysed. The mechanical energy feedback to the ambient ISM can also decrease \logU\ due to the decrease in the density following the feedback-driven expansion of the \hii\ regions. \citet{Popstar2010} found that this effect can decrease \logU\ by as much as  3~dex at the Cartwheel metallicity. All these suggest that the evolution-dependent feedback is driving the observed correlation through the escape of ionizing photons and the decrease in the ISM density. Thus, statistically, the high excitation regions are systematically younger than the low excitation regions.

We also investigated the effect of radiative shocks in \hii\ regions primarily photoionized by clusters as described in Sec.~4.2 above. High velocity shocks tend to increase the line ratios involving low ionization ions such as the \oi\ line \citep[e.g.][]{Stasinska2015}, which would move the points to the right of the ionization sequence in the right panels of Figure~\ref{fig:BPT1}. A tendency of broadening of the sequence in this diagram for low-ionization \hii\ regions is seen, suggesting a possible presence of shocks in some low \logU, which are relatively older, \hii\ regions.

\subsection{The ionizing source of He$^{+}$ in the Cartwheel \hii\ regions}

Having addressed the ionization mechanism and physical processes prevalent in the Cartwheel \hii\ regions, we now investigate whether the same physical mechanisms account for the observed \heiibyhb\ ratio. We have chosen two plots to verify this: \heiibyhb\ vs \ewhb\ and \heiibyhb\ vs \oibyoiii. These are shown in Figure~\ref{fig:heii_mod} for the chosen theoretical tracks from C\&B models, along with special scenarios discussed in the paragraphs above. We do not show the plots with BPASS binary models, as the \heiibyhb\ ratios produced by these models are systematically lower than the observed values as illustrated in Figure~\ref{fig:EWHb_evol2}.

\subsubsection{He$^{++}$ nebulae photoionized by WR stars}

The majority of \heiiwr-emitting regions fall between the tracks corresponding to ionization-bounded \hii\ regions photoionized by clusters of age between 3 and 4~Myr. This age range corresponds to the WR-phase in C\&B models, as can be inferred from Figure~\ref{fig:EWHb_evol2}.

{\it Escape of ionizing photons through holes, and/or the presence of non-ionizing population:} 
The \logU\ values inferred for each region using the \heiibyhb\ vs \ewhb\ are systematically higher by around 1~dex as compared to that inferred from the \heiibyhb\ vs \oibyoiii\ plot. This can be explained as due to the escape of ionizing photons through holes, and/or the presence of a spatially close older cluster in majority of the Cartwheel \hii\ regions, both of which displace the tracks horizontally without changing the flux ratios of the nebular lines. This inference is consistent with the observed correlation between \oiiibyhb\ ratio and \ewhb. 

Is the ionization by WR stars the only way to explain all the observed line ratios. Is the source of ionization of He$^+$ the same as that of other ions? In order to address these questions, we here summarize the results from other scenarios that we have explored. 

{\it Density-bounded models:} The loci of density-bounded \hii\ regions in the line ratio diagrams are shown by green solid lines in the two panels of Figure~\ref{fig:heii_mod} for C\&B models of initial \logU=$-2.5$. The length of the plotted lines correspond to 50~percent of the ionizing photons escaping the nebula from the density bounded zones. The locations of the observed points, with the exception of regions \#144, \#17 and \#148, can be reconciled with this scenario, but for a lower initial \logU\ value, as compared to the ionization-bounded case. However, the observed ionization sequences (see Figure~\ref{fig:oiii_new}) are not consistent with a low value of \logU. Thus, we rule out the possibility that the majority of the Cartwheel \hii\ regions are density-bounded with a low initial \logU.

{\it Role of radiative shocks and two-zone escape models:}
In Figure~\ref{fig:heii_mod}, the radiative shocks and two-zone escape models are plotted for a 3~Myr old ionizing cluster in the right panel, with the aim of covering the observed ratios of regions \#144, \#17 and \#148. It can be inferred from the plot that these scenarios, especially with escape from high \logU\ zone (dashed red line) would cover the observed range of values for the majority of the regions if the ionizing cluster is $\sim$4~Myr old. In fact, we have considered such a possibility to explain the behaviour of points in the right panel of Figure~\ref{fig:heii_mod}.
However, these scenarios produce higher than the observed values of \arivbyariii\ line ratio, as will be discussed later in this section. Thus, we do not find it necessary to look beyond the WR stars to explain the observed He$^{++}$ nebulae. In summary, the majority of the He$^{++}$ nebulae in the Cartwheel are photoionized by WR stars, with the \hii\ regions enclosing the He$^{++}$ nebulae. Different observed quantities can be consistently explained with $\sim$50~percent of the hydrogen ionizing photons escaping through holes, and/or the presence of older non-ionizing populations inside the aperture used for extraction.

\subsubsection{He$^{++}$ nebulae requiring alternative sources of ionization}

Exceptions to the above scenario are five \hii\ regions that standout from the main group. Two (\#99 \& \#112) are at the high-\ewhb\  end, and the other three (\#144, \#17 \& \#148) are among the lowest \ewhb\ \hii\ regions. 

{\it Ionization by ULX source and stripped binary stars:}
The two high-\ewhb\ regions also have the highest \oiiibyhb\ ratio and the lowest \oibyoiii\ ratio, all indicating that these regions have the highest ionization parameter, and the youngest of the sample regions. We infer \logU$\sim-$2.0 and an age corresponding to the pre-WR phase, suggesting that the hot main-sequence stars are the most likely sources of He$^+$ ionization in these two sources.

Region \#99 is the brightest, and the most massive \hii\ region in the Cartwheel with an estimated number of more than 11,000 O stars, assuming O stars are the sole sources of ionization of hydrogen. This region is associated with an ULX source, and hence we discuss here the possible role of this ULX source in the ionization of He$^+$.
The observed \oiiibyhb\ is high enough as expected for the ionization by the hard radiation from a ULX source. However, the observed \oibyha\ ratio for this regions is significantly lower than that expected for the ionization by a ULX source \citep{Gurpide2022}, and hence ULX cannot be the sole or principal source of ionization of this region. The line ratio diagrams presented suggest that the photoionization by the main-sequence stars is a viable source of ionization. The ULX may provide additional photons for the ionization of He$^+$. 
The observed \ewhb, which in spite of being the highest among the sample regions, is still more than a factor of two lower than that expected for single burst of age$<$3~Myr, suggests the presence of underlying non-ionizing populations. It is likely that the star formation in the region is proceeding for more than 3~Myr or that it had a star formation event in the recent past. These slightly evolved stars had enough time to form HMXBs that are generating the X-rays emitted by the ULX source \citep{Plat2019}.

Region \#112 has the lowest value of \heiibyhb\ among the regions studied here, with a value intermediate between that of main-sequence  and WR phases. Like in the case of \#99, all line ratio diagrams presented here suggest ionization of He$^+$ by main sequence stars. 

\begin{figure}
\begin{centering}
\includegraphics[width=1\linewidth]{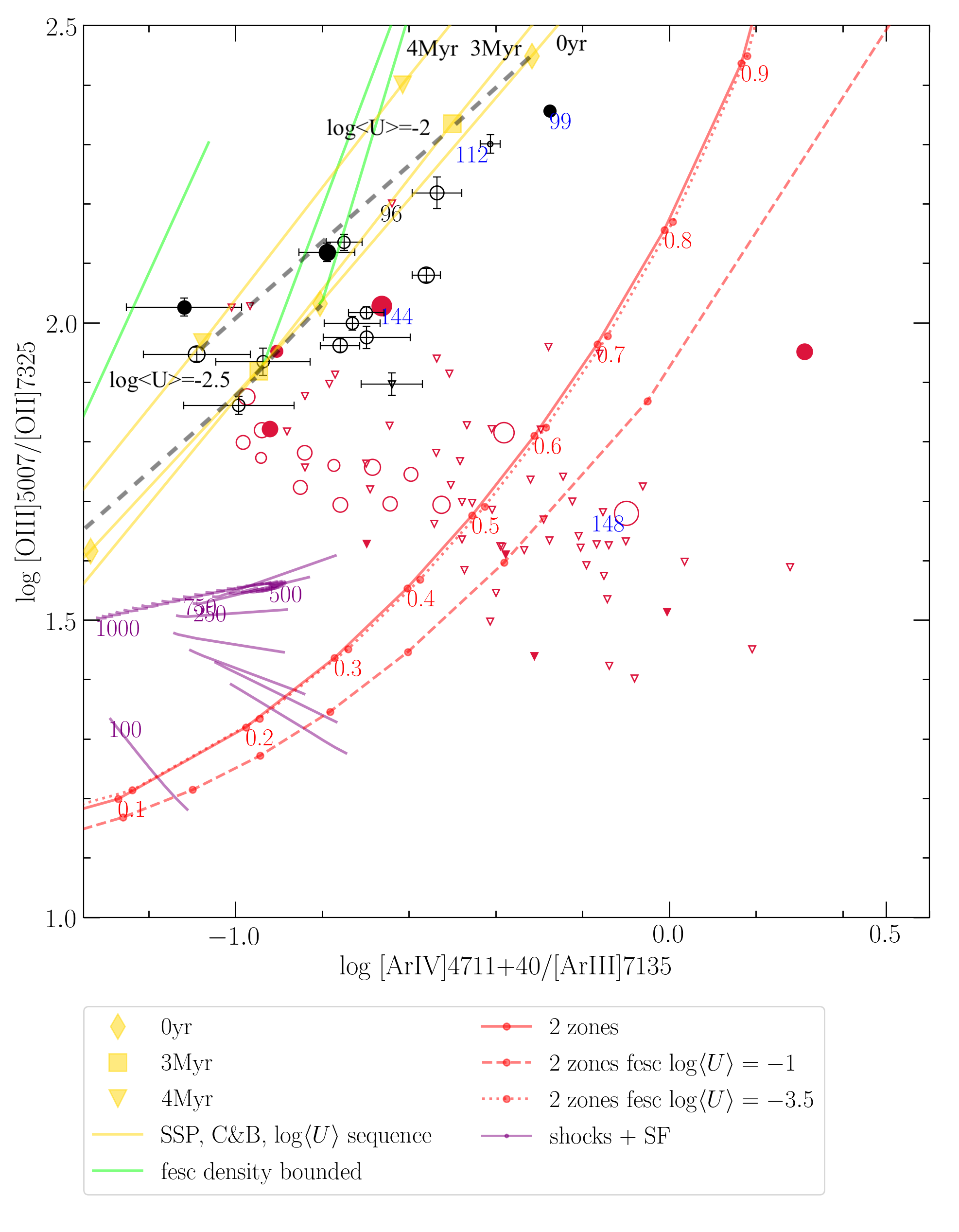}
\par\end{centering}
\caption{
\oiiibyoii\ vs. \arivbyariii\ ratio. \ariv\ lines are detected only in the regions indicated by the black circles, with error bars. All red coloured symbols (circles and triangles) are 3-$\sigma$ upper limits on the \arivbyariii\ ratio. The box below the plot contains a guide to the models plotted. See text for details. 
}
\label{fig:Ar}
\end{figure}

{\it Regions affected by radiative shocks and/or two-zone escape models:}
We now discuss the ionization mechanism of He$^{+}$ in regions \#148, \#17 and \#144, the three low-\ewhb\  regions with the highest ratio of \heiibyhb. These three regions are at the low \logU\  end of the ionization sequence. The \heiibyhb\ ratios predicted by the traditional ionization-bounded case and photoionized by stellar radiation are much lower at these low \logU\ values as compared to the observed values for these three regions. Two of the various processes we have explored produce the observed high \heiibyhb\ ratios at low \logU\ values. These are (1) radiative shock contribution, and (2) two-zone escape models with $\sim$50~percent escape from the high \logU\ zone. Regions \#144, \#17 and \#148, in that order, lie on a sequence of increasing shock velocities with shock velocities in the 100--1000~km\,s$^{-1}$ range, with the \heiibyhb\ value of \#148 only produced by shock models. The two-zone escape models can also explain their location in the plots. Their low \ewhb\ and low \logU\ suggest that these three regions are more evolved than the rest of the regions, and are likely to be in the post-WR phase. We hence favour shock ionization associated with SN explosions as the most likely causes of the observed high values of \heiibyhb\ ratio. SN explosions can also create escape routes for ionizing photons from the high \logU\  zone, and hence shocks and two-zone escape scenarios could be co-existing. Region \#144 is associated with a ULX source whose X-ray luminosity is high enough to contribute to He$^+$ ionization. Thus, X-ray ionization could also be prevalent in \#144.

\citet{Stasinska2015} has advocated the use of the \arivbyariii\ ratios to test the hardness of the ionizing spectrum, especially for the \heiiwr-emitting regions. Ar$^{++}$ has an ionization potential of 40.7~ev and hence the collisionally excited \ariv\ lines are expected to be present in \heiiwr-emitting regions. The \ariv$\lambda\lambda$4711,4740 doublet is relatively faint. Nevertheless, the doublet is detected at more than 3$\sigma$ levels in 14 of the 32 \heiiwr-emitting regions. In Figure~\ref{fig:Ar}, we plot the \oiiibyoii\ ratios against the \arivbyariii\ line ratio. For photoionized nebulae, the locus of points in this diagram is mainly governed by \logU, with the observed points lying between $-3.0<$\logU$<-2.0$. Spectral evolution in the age range 0--4~Myr introduces small spread in the direction perpendicular to the sequence formed by ranges of initial \logU. On the other hand, the presence of radiative shocks and/or selective escape of photons from low or high \logU\ zones moves the points to the right (i.e. higher \arivbyariii\ ratios). Thus, this figure is useful to discriminate the purely photoionized models from the photoionized+shock models, or the models involving selective escape of photons. Unfortunately the last two cases follow similar trajectories in the diagram, and hence cannot be distinguished.

The main-group of the \heiiwr-emitting regions in which \ariv\ lines have been detected lie along the  photoionization sequence by the SSP models, and more importantly, are not consistent with the presence of shocks and/or two-zone escape scenarios, independent of the age of the ionizing clusters. Thus, this diagram helps us to break the degeneracy seen in Figure~\ref{fig:heii_mod}. Region \#99 lies clearly to the right of the photoionization sequence, reiterating that the hard radiation from the ULX source plays a role in the ionization of ions that have higher than 40~eV of ionization potential. The second region in which we cannot rule out  ionization by the ULX source is \#144. The \ariv\ lines are not detected in this region, but the upper limit on the \arivbyariii\ ratio is slightly to the right of the photoionization sequence by the SSP models. Unfortunately, we have only upper limits on the detection of the \ariv\ lines in the remaining two interesting regions \#148 and \#17. The observed upper limits in \#148 and \#17 are consistent with the presence of shocks and/or two-zone escape scenarios.

\subsection{Regions with non-detection of \heii$\lambda$ 4686 line}

We here carry out an analysis on the nature of the \hii\ regions where the \heiiwr\ line could not be detected at the 3-$\sigma$ confidence level. 
Given that the WR stars are the principal sources of He$^+$ ionizing photons in our sample regions, and that the WR stars appear only for a short duration in an SSP, this fraction is expected to be a function of age. The \ewhb\ is an excellent proxy for age in young stellar systems (see Figures~\ref{fig:EWHb_evol1}), and hence we analyse the detection fraction as a function of \ewhb. From the analysis of Figures~\ref{fig:oiii_new} and \ref{fig:heii_mod}, we arrived at the conclusion that quantitatively the observed values of \ewhb\ are systematically smaller as compared to the values expected for ionization-bounded \hii\ regions using C\&B models. Continuum contribution from a non-ionizing population and the escape of ionizing photons from the nebula are two principal mechanisms that lead to a decrease in \ewhb\ from those expected for ionization-bounded \hii\ regions ionized by a single-age population. The factor by which the \ewhb\ is reduced may vary from region to region. For the sake of using \ewhb\ as a proxy for age, we assume a reduction factor anywhere between 0 and 50\% for the sample regions.

\begin{figure}
\begin{centering}
\includegraphics[width=1\linewidth]{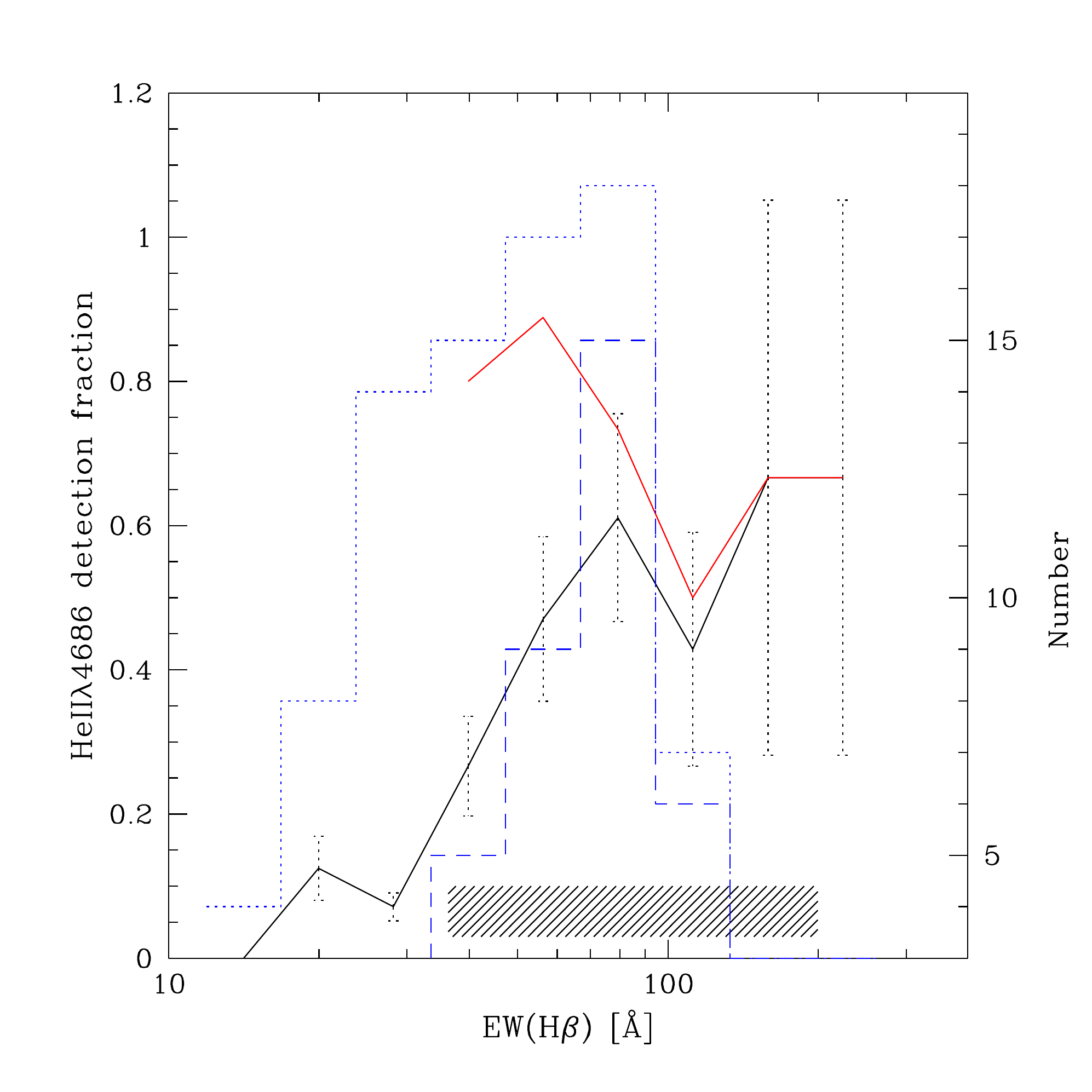}
\par\end{centering}
\caption{
Fraction of \hii\ regions with detection of the \heiiwr-line as a function of the \ewhb\ for the whole sample (black solid line) and a sample of \hii\ regions that had sensitivity to detect the \heiiwr\ line if \heiibyhb$\geq$0.01 (red solid line). The distribution of the \ewhb\ for the whole sample (dotted histogram) and a sample of \hii\ regions that had sensitivity to detect the \heiiwr\ line if \heiibyhb$\geq$0.01 (dashed histogram) are also plotted, with the right axis showing the numbers. For reference, we also show the range of \ewhb\ when the number of WR stars in the C\&B 
$M_{\rm u}=100$~\msol\ models is non-zero, for an ionization-bounded \hii\ region (the shaded area). The error on the derived fractions for the whole sample is shown, which is based on the square-root of the number of \hii\ regions in each bin (the sizes of the error bars on the red line are similar and hence we omit them for the sake of clarity of the figure).
}
\label{fig:detection_fraction}
\end{figure}

In Figure~\ref{fig:detection_fraction}, we show the fraction of \hii\ regions detected as a function of the observed \ewhb. The theoretically expected range of \ewhb\ during the WR phase is shown by the shaded area, which takes into account the reduction of \ewhb\ by  0 to 50\%  during the WR phase. The distribution of observed \ewhb\ for the whole sample (dotted histogram), and the sample of regions where we have achieved a 3-$\sigma$ sensitivity to detect the \heiiwr\ line if they had \heiibyhb$\geq$0.01 (dashed histogram), are shown. For the whole sample (black line), the detection fraction decreases with decreasing \ewhb. For the subset of \hii\ regions (red line) that have SNRs sufficient to detect the \heiiwr\ line for typical values during WR phase (\heiibyhb$\geq$0.01), the detection fraction peaks at \ewhb$\sim$60~\AA. The peak value reaches as high as 90\%. The \ewhb\ at the peak value corresponds to that during the WR phase. This suggests that our dataset is sensitive enough to detect almost all ($\sim$90\%) He$^{++}$ nebulae ionized by the WR stars. Before the onset of the WR phase (at the high \ewhb-end), the detection fraction is nonzero. The \heiiwr\ line is detected in two of the three \hii\ regions, indicating that the \heiibyhb\ ratio might be higher than the values predicted in the current SSPs during the main sequence phase of stars. The ULX is the likely source of additional ionization in one of these (\#99), whereas stripped binary stars, which are not included in the C\&B models, could be the possible source of the weak ionization in the other (\#112).

The rest of the non-detections (26) corresponds to low-\ewhb\ regions. In the C\&B models, these regions correspond to the post-WR phase. However, the inclusion of the binary channel for the formation of WR stars in the BPASS models extends the duration of WR phase to these low \ewhb\ values, with the expected ratio of \heiibyhb\ lower than that during the WR phase in C\&B models. Unfortunately, we do not reach the sensitivity to detect the \heiiwr\ line with \heiibyhb$<$0.01 for regions with \ewhb$<40$~\AA\ regions, and hence the non-detection of the \heiiwr\ line could be due to the absence of He$^{++}$ ions in these regions, or that we do not reach the sensitivity level required to detect weak ionization from the WR stars formed through the binary channel.

In general, the \hii\ regions with \heiiwr-line detection have higher ionization parameter as compared to the \hii\ regions without the \heiiwr-line detection, at each \ewhb\ bin. 

The location of \hii\ regions with non-detection of \heiiwr\ line in Figure~\ref{fig:Ar} suggests shock and/or two-zone escape scenarios are more prevalent in these regions as compared to the \heiiwr-emitting main-group \hii\ regions. These processes are strong enough to increase the \heiiwr\ lines above the detectable limits in only three cases (\#144, \#17 and \#148).

\section{Conclusions}

We have carried out a search for \heiiwr\ nebular line in the Cartwheel \hii\ regions using the VLT/MUSE datacube. We detect the \heiiwr\ line in 32 \hii\ regions, with a mean value of \heiibyhb=0.010$\pm$0.003. All the detections are situated in the star-forming ring of the Cartwheel, with ten of these sources coinciding with the location of a ULX source.
We use commonly used diagnostic line ratios to compare the ionization properties of \hii\ regions with and without the detection of the \heiiwr\ line. The \heiiwr\ line-emitting regions with and without the ULX sources, in general, show similar ionization properties in the diagnostic diagrams. Hence, the ULX sources are not the principal suppliers of ionizing photons in all the \hii\ regions containing ULX sources. 
Analysis of the diagnostic diagrams using C\&B SSPs suggests that the majority (27) of the detections correspond to \hii\ regions in their WR phase, with two and three detections corresponding to \hii\ regions in their pre-WR and post-WR phases, respectively. 
However, the characteristic BB indicating the presence of WR stars is not detected in our sample regions. We illustrate that this non-detection is due to the relatively low EWs of the BB in SSPs for IMFs with $M_{\rm u}\leq$100~\msol\ at the metallicity of the Cartwheel, even when the SSPs have sufficient number of WR stars to provide the ionization of He$^+$. We suggest that main sequence stars are the major contributors to ionization in the two pre-WR \hii\ regions, with an additional contribution from other hard sources. In region\#99, this additional contribution most likely comes from the ULX source. On the other hand, the three \hii\ regions in the post-WR phase may be either ionized by radiative shocks or their \hii\ regions are leaky. We find a correlation between \oiiibyhb\ and \ewhb, which requires a more rapid softening of the ionization parameter \logU\ than that considered in C\&B SSP models. This rapid softening can be naturally explained if the \hii\ regions expand as the cluster evolves due to the feedback from massive stars. 
The detection frequency of the \heiiwr\ line reaches values as high as 90\% for \hii\ regions that have \ewhb=40--70~\AA. These values of \ewhb\  correspond to late stages of the WR phase in the C\&B models. Our dataset lacks sensitivity to detect the \heiiwr\ line from \hii\ regions with \ewhb$<$40~\AA, when WR stars formed from the binary channel are expected to dominate the ionization of He$^+$. 

\section*{Acknowledgements}

We thank an anonymous referee for many thoughtful comments that improved the paper. We also thank Gerardo Ramos-Larios who helped us in preparing the images appearing in Figures~\ref{fig:muse_image} and \ref{fig:zoom_image}, and CONACyT for the research grant CB-A1-S-25070 (YDM).
This work is based on data obtained from the ESO Science Archive Facility, program ID: 60.A-9333. Observations made with the NASA/ESA {\it Hubble Space Telescope} were obtained from the data archive at the Space Telescope Science Institute. STScI is operated by the Association of Universities for Research in Astronomy, Inc. under NASA contract NAS 5-26555.

\section*{Data availability}

The fluxes of principal emission lines used in this work are available
in the article and in its online supplementary material. The reduced fits files
on which these data are based will be shared on reasonable request to the first
author.
The ESO datacubes are in the public domain. The link and observation ID are available in the article.



\begin{thebibliography}{99}

\bibitem[\protect\citeauthoryear{Alarie \& Morisset}{2019}]{Alarie2019} Alarie A., Morisset C. 2019, RMxAA, 55, 377 

\bibitem[\protect\citeauthoryear{Appleton \& Struck-Marcel}{1996}]{1996FCPh...16..111A}
Appleton, P.~N. \& Struck-Marcel, C. 2010, \fcp, 16, 111 

\bibitem[\protect\citeauthoryear{Baldwin, Phillips, \& Terlevich}{1981}]{BPT1981} 
Baldwin J.~A., Phillips M.~M., Terlevich R. 1981, PASP, 93, 5


\bibitem[\protect\citeauthoryear{Bacon et al.}{2010}]{Bacon2010}
Bacon, R. et al.\  2010, \procspie, 7735, 773508

\bibitem[\protect\citeauthoryear{Belfiore et al.}{2022}]{Belfiore2022}
Belfiore, F., Sontoro, F., Groves, B. et al. 2022, A\&A, 659, 26

\bibitem[\protect\citeauthoryear{Bottorff et al.}{1998}]{Bottorff1998} 
Bottorff M., Lamothe J., Momjian E., Verner E., Vinkovi{\'c} D., Ferland G., 1998, PASP, 110, 1040 

\bibitem[\protect\citeauthoryear{Bressan et al.}{2012}]{Bressan2012}
Bressan, A., Marigo, P., Girardi, L. et al. 2012, MNRAS, 427, 127




\bibitem[\protect\citeauthoryear{Bruzual \& Charlot}{2003}]{Bruzual2003}
Bruzual, A.~G., Charlot, S. 2003, \mnras, 344, 1000


\bibitem[\protect\citeauthoryear{Castellanos, Di\'az \& Tenorio-Tagle}{2002}]{Castellanos2002} 
Castellanos, M., Di\'az, A. I., Tenorio-Tagle, G. 2002, \apj, 579, L79


\bibitem[\protect\citeauthoryear{Chabrier}{2003}]{Chabrier2003}
Chabrier, G. 2003, PASP, 115, 763

\bibitem[\protect\citeauthoryear{Charlot \& Fall}{1993}]{Charlot1993} Charlot S., Fall S.~M. 1993, ApJ, 415, 580

\bibitem[\protect\citeauthoryear{Chen et al.}{2015}]{Chen2015}
Chen, Y., Bressan, A., Girardi, L., et al.\ 2015, MNRAS 452, 1068


\bibitem[Crowther(2007)]{2007Crowther} Crowther, P.~A.\ 2007, \araa, 45, 177

\bibitem[\protect\citeauthoryear{Eldridge et al.}{2017}]{Eldridge2017}
Eldridge, J.~J., Stanway, E.~R., Xiao, L. et al. 2017, PASA, 34, e058

\bibitem[\protect\citeauthoryear{Erb et al.}{2010}]{Erb2010} 
Erb, D. K., Pettini M., Shapley A. E., Steidel C. C.. Law D. R., Reddy N. A., 2010, ApJ, 719, 2 


\bibitem[\protect\citeauthoryear{Feltre, Charlot, \& Gutkin}{2016}]{Feltre2016} 
Feltre A., Charlot S., Gutkin J. 2016, MNRAS, 456, 3354 

\bibitem[\protect\citeauthoryear{Ferland et al.}{2017}]{Ferland2017} Ferland G.~J., Chatzikos M., Guzm{\'a}n F., et al.
2017, RMxAA, 53, 385

\bibitem[\protect\citeauthoryear{Fosbury \& Hawarden}{1977}]{Fosbury1977}
Fosbury, R. A. E. \& Hawarden, T. G. 1977, MNRAS, 178, 473 




\bibitem[\protect\citeauthoryear{Gao et al.}{2003}]{Gao2003}
Gao, Yu, Wang, Q. Daniel, Appleton, P. N. \& Lucas, Ray A. 2003, ApJ, 596, L171 

\bibitem[\protect\citeauthoryear{Gil de Paz et al.}{2018}]{2018SPIE10702E..17G}
Gil de Paz, A., Carrasco, E., Gallego, J. et al. 2018, SPIE,  1070217, SPIE10702


\bibitem[\protect\citeauthoryear{G{\"o}tberg et al.}{2018}]{Gotberg2018} 
G{\"o}tberg Y., de Mink S.~E., Groh J.~H., Kupfer T., Crowther P.~A., Zapartas E., Renzo M. 2018, A\&A, 615, A78. 

\bibitem[\protect\citeauthoryear{G\'omez-Gonz\'alez et al.}{2020}]{Gomez-Gonzalez2020}
G\'omez-Gonz\'alez, V.M.A., et al. 2020, \mnras, 493, 3879

\bibitem[\protect\citeauthoryear{G\'omez-Gonz\'alez et al.}{2021}]{Gomez-Gonzalez2021}
G\'omez-Gonz\'alez, V.M.A., et al. 2021, \mnras, 500, 2076

\bibitem[\protect\citeauthoryear{G\'urpide et al.}{2022}]{Gurpide2022}
G{\'u}rpide A., Parra M., Godet O., Contini T., Olive J.-F., 2022, A\&A, 666, A100    

\bibitem[\protect\citeauthoryear{Gutkin et al.}{2016}]{Gutkin2016}
Gutkin J., Charlot S. \& Bruzual, G. 2016, \mnras, 462, 1757

\bibitem[\protect\citeauthoryear{Higdon}{1995}]{Higdon1995}
Higdon, J. L. 1995, ApJ, 455, 524 





\bibitem[\protect\citeauthoryear{Kehrig et al.}{2015}]{Kehrig2015}
Kehrig, C., V\'{\i}lchez, J.~M., P\'erez-Montero, E. et al. 2015, \apjl, 80, L28

\bibitem[\protect\citeauthoryear{Kehrig et al.}{2018}]{Kehrig2018}
Kehrig, C., V\'{\i}lchez, J.~M., Guerrero, M.~A. et al. 2018, \mnras, 480, 1081


\bibitem[\protect\citeauthoryear{Kim, Kim \& Ostriker}{2019}]{Kim2019}
Kim, Jeong-Gyu, Kim, Woong-Tae, Ostriker, Eve C. 2019, \apj, 883, 102






\bibitem[\protect\citeauthoryear{L{\'o}pez-S{\'a}nchez \& Esteban}{2010}]{Esteban2010}
L{\'o}pez-S{\'a}nchez {\'A}. R., Esteban C. 2010, A\&A, 516, A104


\bibitem[\protect\citeauthoryear{Lynds \& Toomre}{1976}]{1976ApJ...209..382L}
Lynds, R. \& Toomre, A. 1976, \apj, 209, 382


\bibitem[\protect\citeauthoryear{Marcum, Appleton \& Higdon}{1992}]{Marcum1992}
Marcum, P.M, Appleton, P.N, \& Higdon, J. L. 1992, ApJ, 399, 57 

\bibitem[\protect\citeauthoryear{Mart{\'\i}n-Manj{\'o}n et al.}{2010}]{Popstar2010}
Martín-Manjón, M. L., García-Vargas, M. L., Mollá, M., Díaz, A. I. 2010, \mnras, 403, 2012

\bibitem[\protect\citeauthoryear{Mapelli et al.}{2010}]{Mapelli2010}
Mapelli, M., Ripamonti, E., Zampieri, L., Colpi, M. \& Bressan, A. 2010, MNRAS, 408, 234


\bibitem[\protect\citeauthoryear{Mayya \& Prabhu}{1996}]{Mayya1996} 
Mayya Y.~D., Prabhu T.~P. 1996, AJ, 111, 1252  

\bibitem[\protect\citeauthoryear{Mayya et al.}{2020}]{Mayya2020}
Mayya, Y. D., Carrasco, E., G\'omez-Gonz\'alez, V. M. A. et al. 2020, MNRAS, 498, 1496





\bibitem[\protect\citeauthoryear{Osterbrock \& Ferland}{2006}]{Osterbrock2006}
Osterbrock, D.~E. \& Ferland, G.~J. 2006, in Astrophysics of Gaseous Nebulae and Active Galactic Nuclei (CA: University Science Books)



\bibitem[\protect\citeauthoryear{P\'erez-Montero}{2017}]{Perez-Montero2017}
P\'erez-Montero, E. 2017, PASP, 129, 043001

\bibitem[\protect\citeauthoryear{Plat et al.}{2019}]{Plat2019}
Plat, A., Charlot, S., Bruzual, G., et al.
2019, MNRAS, 490, 978


\bibitem[\protect\citeauthoryear{Ramambason et al.}{2020}]{Ramambason2020}
Ramambason L., Schaerer D., Stasi{\'n}ska G., et al.,
2020, A\&A, 644, A21 

\bibitem[\protect\citeauthoryear{Renaud et al.}{2018}]{Renaud2018} 
Renaud F., Athanassoula E., Amram P., et al. 2018, MNRAS, 473, 585

\bibitem[\protect\citeauthoryear{Schaerer}{1996}]{Schaerer1996}
Schaerer, D., 1996, \apj, 467, L17

\bibitem[\protect\citeauthoryear{Schaerer, Fragos \& Izotov}{2019}]{Schaerer2019}
Schaerer D., Fragos T., Izotov Y.~I., 2019 A\&A, 622, L10



\bibitem[\protect\citeauthoryear{Shirazi \& Brinchmann}{2012}]{Shirazi2012}
Shirazi, M. \& Brinchmann, J. 2012, \mnras, 421, 1043


\bibitem[\protect\citeauthoryear {Stasinska \& Leitherer}{1996}]{Stasinska1996}
Stasińska G., Leitherer, C. 1996, ApJSS, 107, 661

\bibitem[\protect\citeauthoryear {Stasinska et al.}{2015}]{Stasinska2015}
Stasińska G., Izotov Yu., Morisset, C., Guseva, N. 2015, A\&A, 576, A83

\bibitem[\protect\citeauthoryear {{Stanway} \& {Eldridge}}{2018}]{Stanway2018}
Stanway E. R., Eldridge J. J. 2018, \mnras, 479, 75

\bibitem[\protect\citeauthoryear{Struck}{2010}]{2010MNRAS.403.1516S}
Struck, C. 2010 \mnras, 403, 1516 

\bibitem[\protect\citeauthoryear{Todt et al.}{2015}]{Todt2015}
Todt, H., Sander, A., Hainich, R. et al. 2015, A\&A, 579, A75

\bibitem[\protect\citeauthoryear{Tresse et al.}{1999}]{Tresse1999}
Tresse, L., Maddox, S., Loveday, J. \& Singleton, C. 1999, \mnras, 310, 262


\bibitem[\protect\citeauthoryear{Wolter \& Trinchieri}{2004}]{Wolter2004}
Wolter, A. \& Trinchieri, G. 2004, A\&A, 426, 787

\bibitem[\protect\citeauthoryear{Wolter, Fruscione \& Mapelli}{2018}]{Wolter2018}
Wolter, A., Fruscione, A. \& Mapelli, M. 2018, ApJ, 863, 43


\bibitem[\protect\citeauthoryear{Zaragoza-Cardiel et al.}{2022}]{Zaragoza2022}
Zaragoza-Cardiel J., G{\'o}mez-Gonz{\'a}lez V.~M.~A., Mayya D., Ramos-Larios G., 2022, MNRAS, 514, 1689 

\end{thebibliography}
\end{document}